\documentclass{aa}  

\usepackage{graphicx}
\usepackage{txfonts}
\usepackage{placeins}
\usepackage{float}
\newcommand{\kms}{km\,s$^{-1}$}
\newcommand{\vr}{$V_{\rm r}$}

\newcommand{\porb}{$P_{\rm orb}$}
\newcommand{\vsini}{$v\sin i$}

\begin{document}

   \title{Dissimilar magnetically driven accretion on the components of V4046 Sagittarii}

   \author{K. Pouilly\inst{1}
          \and
          M. Audard\inst{1}
          }

   \institute{Department of Astronomy, University of Geneva, Chemin Pegasi 51, CH-1290 Versoix, Switzerland\\
              \email{Kim.Pouilly@unige.ch}
             }

   \date{Received 4 February 2025; Accepted 8 April 2025}

 
  \abstract
   {Accretion of pre-main sequence stars (PMS) is a key process in stellar formation, governing mass assembly, influencing angular momentum conservation and stellar internal structure, and shaping disc evolution, which serves as the birthplace of exoplanets. Classical T Tauri stars (cTTSs), low-mass PMS stars actively accreting from a disc, hold a well-described magnetospheric accretion model. Their strong, inclined dipole magnetic fields truncate the disc at a few stellar radii, channelling material along magnetic field lines to fall onto the stellar surface near the dipole pole. However, this paradigm assumes the presence of a single star, and a complete description of the accretion process in multiple systems remains to be achieved.}
   {Building on our previous work on DQ Tau and AK Sco, we aim to describe the accretion processes in cTTS binaries, accounting for the influence of stellar magnetic fields. Specifically, we sought to explore how the magnetospheric accretion model of cTTSs can be applied to V4046 Sgr, a spectroscopic binary composed of equal-mass and coeval cTTSs in a circular orbit with synchronous rotation, surrounded by a circumbinary disc.}
   {We analysed a time series of ESPaDOnS spectra covering several orbital cycles. A variability analysis was performed on the radial velocities and on the Balmer, \ion{He}{I} D3, and \ion{Ca}{II} emission lines, which are associated with the accretion process.}
   {We identified the secondary as the system's main accretor, operating in an unstable regime. Additionally, we detected an accretion funnel flow connecting the dipole pole of the primary star with a nearby bulk of gas.}
   {We concluded that the two components exhibit dissimilar accretion patterns. The primary operates in an "ordered chaotic" regime, where accretion funnel flows and accretion tongues (which penetrate the magnetosphere to reach the stellar equator) coexist. Conversely, the secondary appears to be in a chaotic regime, with accretion tongues dominating.}

   \keywords{Stars: variables: T Tauri --
                Stars: individual: V4046 Sgr --
                Accretion, accretion disks --
                Techniques: spectroscopic  
               }

   \maketitle
%

\section{Introduction}

    The accretion process in classical T Tauri stars (cTTSs)—young, low-mass stellar objects still surrounded by a disc from which they accrete material—is commonly described by the magnetospheric accretion model \citep{Hartmann94}.
    In this model, the strong dipolar magnetic field of such systems exerts magnetic pressure on the disc at the magnetospheric radius, forcing material to leave the disc plane and fall onto the stellar surface along magnetic field lines. 
    While simulations have successfully reproduced this framework \citep[see the review by][]{Romanova15}, and its characteristics are often observed in cTTSs, a major assumption underlying the model is that the system must be single.
    Given that most low-mass stars form in multiple systems \citep{Offner23}, this accretion model applies to only a minority of stars.
    Therefore, dedicated studies are required to develop a framework that can accommodate multiple systems.
    This work extends previous efforts in this direction, such as studies of the eccentric cTTS binaries DQ Tau and AK Sco \cite{Pouilly23, Pouilly24}, by investigating systems with different orbital configurations, starting with the present study on \object{V4046 Sagittarii}.

    \object{V4046 Sgr} is a type II spectral binary (SB2) system \citep[possibly part of a hierarchical quadruple system, with the binary GSC0739 as a distant--$\sim$12\,350 au--companion;][]{Kastner11} comprising two cTTSs of spectral types K5 and K7.
    The system is surrounded by a circumbinary disc with an inner radius of 0.35 au.
    The two stars orbit in a circularised configuration with a projected separation of approximately 2.5 R$_\odot$.
    Their rotational periods are synchronised with the 2.42-day orbital period, and the rotation axes are aligned with both the orbital motion and the disc, which is observed at an inclination of 35$^\circ$.
    The disc's mass is estimated to be between 0.01 and 0.08 M$_\odot$, with its thickness increasing to 0.2 au at the outer edge, located approximately 100 au from the system's centre \citep{Quast00}.

    Further characterisation of the large-scale circumbinary disc's structure was provided by \cite{Martinez22}, who used ALMA 1.3 mm continuum imaging, SPHERE-IRDIS polarised images, and a well-sampled spectral energy distribution. 
    Their model predicts the presence of an inner disc at 5 au, but too faint to be detected by ALMA.
    Nevertheless, they detected a narrow ring at 13 au with a width of 2.46 au and a height 10 times lower, a 10 au gap, and another ring at 24 au.
    The latter exhibits peak intensity at 30 au and a break in luminosity at 36 au.

    The orbital parameters of V4046 Sgr were refined by \cite{Stempels04} through a series of UVES spectra.
    Their analysis yielded a more consistent mass ratio for the components, confirmed the circularisation of the orbit, and verified the synchronous rotation.
    Notably, they determined an orbital period of 2.4213459 days, with semi-amplitudes K$_{\rm 1}$=54.16~\kms\ and K$_{\rm 2}$=56.61~\kms.
    Spectral line analysis revealed low levels of veiling ($\sim$5\% of the continuum at 620 nm) and minimal or absent extinction.
    Examination of emission lines indicated a \ion{Ca}{ii} K line with a narrow component (NC) for both stars, closely tied to the binary orbit, suggesting formation in extended chromospheres around each star.
    Additionally, they decomposed higher-order Balmer lines (H8, H9, H10), which shared similar shapes, into four components: an NC (close to the \ion{Ca}{ii} K NC) and a broad component (BC) for both stars.
    The BC varied with the orbital period, showing an amplitude of 80~\kms, larger than the stellar orbital amplitude.
    They concluded that these BC emissions originate from gas bulks co-rotating at 6.9 R$_\odot$ from the centre of mass, corresponding to approximately 1.15 R$_\star$ from each star.
    This is well within the inner edge of the circumbinary disc and near the co-linear Lagrange points. The stars appear to accrete from these bulks, with the BC's variable equivalent width (EW) indicating variable accretion rates.

    In 2009, V4046 Sgr was the subject of a multi-instrument campaign that included X-ray observations with XMM-Newton \citep{Argiroffi12}.
    These observations revealed the periodic signature of an accretion shock at half the orbital period.
    Three hypotheses were proposed to explain this signal: (i) one component has two accretion shock regions at opposite longitudes, (ii) one component has a single accretion region that emits maximally when the shock is viewed edge-on, or (iii) both components have symmetric shocks located 180$^\circ$ apart relative to the binary rotation axis.

    The campaign also included quasi-simultaneous optical observations with the Echelle SpectroPolarimetric Device for Observation of Stars \citep[ESPaDOnS;][]{Donati03} mounted on the Canada-France-Hawaii Telescope (CFHT).
    \cite{Donati11b} analysed these data to study the system's large-scale magnetic field.
    The magnetic topologies of the two components were found to be complex and weak, reflecting the partly convective internal structure of these 12 Myr-old stars.
    The primary exhibited a mean field strength ($\langle$B$\rangle$=230 G) with a dominant azimuthal component and a maximum field strength (B$_{\rm max}$) of 500 G in an arc-like structure near the pole.
    Its dipole component was weak (B$_{\rm dip}$=100 G), non-axisymmetric, and had an obliquity of 60$^\circ$, facing the observer at phase 0.8.
    The secondary had a weaker mean field strength (B$_{\rm max}$=170 G) and a highly oblique dipole (B$_{\rm dip}$=70 G, obliquity ~90°), facing the observer at phase 0.1.
    Surface maps revealed cool spots near the rotational poles of both stars, slightly offset towards the hemispheres facing their companion.
    Low-contrast, extended Ca II infrared triplet (IRT) excess emissions suggested accretion occurs at multiple sites on the stellar surface, rather than concentrating in a specific region.

    \cite{Hahlin22} also used the ESPaDOnS data set to study the small-scale magnetic field through Zeeman intensification of a Ti multiplet.
    They found similar small-scale field strengths for both components (1.96 and 1.83 kG for the primary and secondary, respectively) and nearly identical filling factors.
    Their analysis yielded a smaller luminosity ratio ($LR$ = 1.27) than that determined by \cite{Stempels04} and higher \vsini\ values (15.1 and 14.4 \kms) for the primary and secondary, respectively.

    In this paper, we provide a detailed description of the system's accretion process using the same data set employed for magnetic field studies. 
    Observations are described in Sect.~\ref{sec:obs}, analyses and results are presented in Sect.~\ref{sec:results}, and these findings are discussed in Sect.~\ref{sec:discussion}. We conclude this work in Sect.~\ref{sec:conclusion}.

\section{Observations}
\label{sec:obs}

    The dataset used in this work is the same as that utilised by \cite{Donati11b} and \cite{Hahlin22} for their magnetic analysis of V4046~Sgr.
    It consists of eight spectropolarimetric observations acquired at the CFHT using ESPaDOnS.

    This instrument covers a wavelength range of 370--1050 nm, achieving a resolving power of 68\,000. It was operated in spectropolarimetric mode, meaning that each observation consists of four sub-exposures taken with different polarimeter configurations. 
    The observations were reduced using the automatic pipeline \texttt{Libre-ESpRIT}, which combines the sub-exposures to optimise the extraction of ESPaDOnS' unpolarised (Stokes \textit{I}) and circularly polarised (Stokes \textit{V}) spectra.

    In this work, only the Stokes \textit{I} spectra are used, as the Stokes \textit{V} spectra were fully analysed in \cite{Donati11b}, but we included in Sect.~\ref{sec:discussion} a discussion about the overall Stokes \textit{I} and \textit{V} results.
    The observations were carried out between 3 September 2009 and 9 September 2009, following a 1-day cadence (except for 6 September 2009, when the star was observed twice).
    The signal-to-noise ratios (S/N) per spectral pixel at order 31 (730 nm) range between 140 and 179.
    A journal of observations is provided in Table~\ref{tab:obs}.
    
    \begin{table}
    \centering
    \caption{Log of ESPaDOnS observations used in this work.}
    \begin{tabular}{llll}
        \hline
        Date & HJD & S/N$_{\rm I}$ & S/N$_{\rm LSD}$ \\
        (2009) & ($-$2\,450\,000 d) & & \\
        \hline
        03 Sept & 5077.77441 & 147 & 1894 \\
        04 Sept & 5078.81732 & 140 & 1512 \\
        05 Sept & 5079.72223 & 168 & 1906 \\
        06 Sept & 5080.72214 & 174 & 2041 \\
        06 Sept & 5080.81413 & 179 & 2009 \\
        07 Sept & 5081.72205 & 142 & 1968 \\
        08 Sept & 5082.72296 & 178 & 1924 \\
        09 Sept & 5083.72186 & 184 & 1674 \\
        \hline  
    \end{tabular}
    \tablefoot{S/N$_{\rm I}$ is the peak S/N by spectral pixel at order 31 (730 nm) and S/N$_{\rm LSD}$ corresponds to the effective S/N of the LSD Stokes \textit{I} profiles (see Sect.~\ref{subsec:LSDVr}).}
    \label{tab:obs}
    \end{table}

\section{Results}
\label{sec:results}
    
    \subsection{Least-Square Deconvolution profiles and radial velocities}
    \label{subsec:LSDVr}
 
    To study the orbital modulation through the radial velocity (\vr) variation, we utilised the Least Squares Deconvolution \citep[LSD;][]{Donati97} Stokes \textit{I} profiles, which represent a weighted average of as many photospheric lines as possible.
    These profiles were computed using the \texttt{LSDpy}\footnote{\url{https://github.com/folsomcp/LSDpy}} Python package.

\begin{figure}
        \centering
        \includegraphics[width=0.99\linewidth]{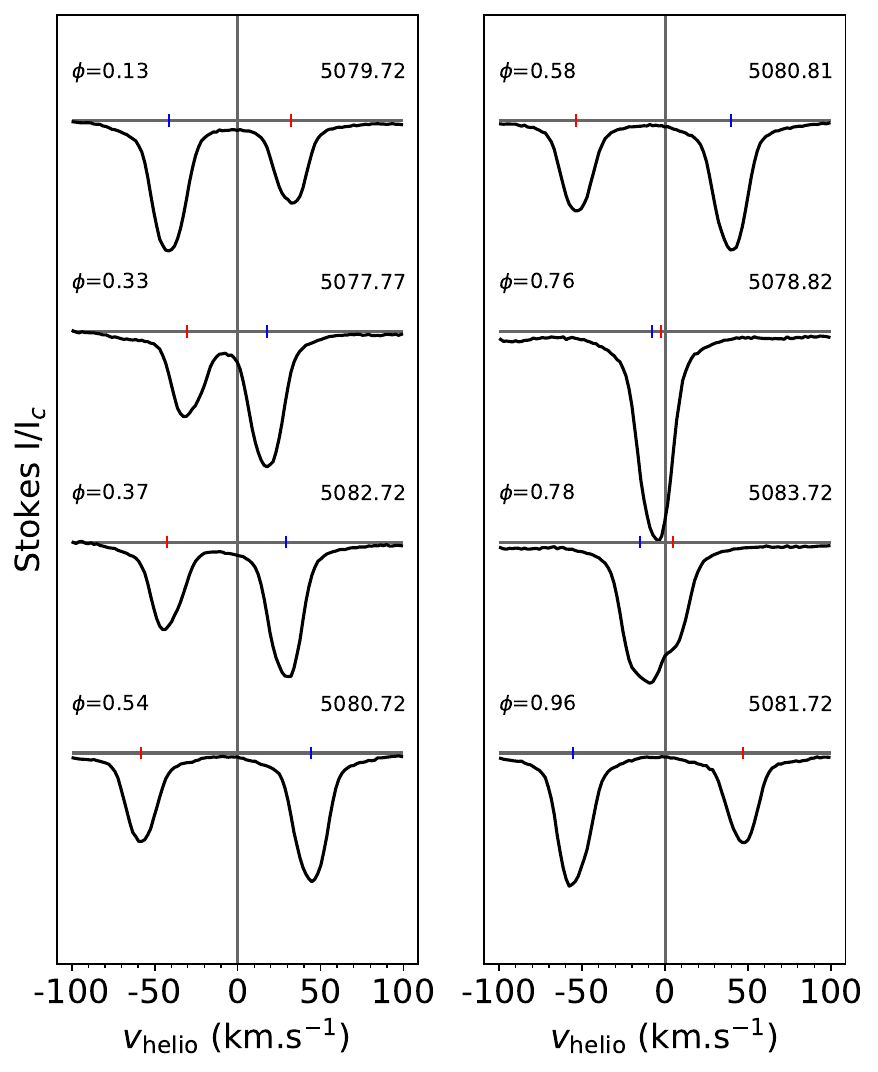}
        \caption{LSD Stokes \textit{I} profiles of V4046~Sgr. The blue (red) ticks illustrate the velocity of the A (B) component. The orbital phases (computed from Eq.~\ref{eq:ephemeris}) and HJDs ($-$2\,450\,000 d) are indicated on the left and right of each profile, respectively.}
        \label{fig:stokesI}
    \end{figure}

    The LSD weights were normalised using an intrinsic line depth of 0.2, a mean Landé factor of 1.310, and a mean wavelength of 500 nm.
    Photospheric lines were selected by creating a mask based on the \texttt{VALD} database line list \citep{Ryabchikova15}, with \texttt{MARC} \citep{Gustafsson08} atmospheric models tailored to the parameters of V4046~Sgr.
    We then removed emission lines and heavily blended lines using the \texttt{SpecpolFlow}\footnote{\url{https://github.com/folsomcp/specpolFlow}} Python package, resulting in approximately 15\,000 lines used to compute the LSD profiles.
    The resulting profiles are presented in Fig.~\ref{fig:stokesI}, with their S/N ranging from 1512 to 2041.

    To derive the \vr, we employed the disentangling procedure described in \cite{Pouilly23}.
    This method was originally designed to determine a mean disentangled profile for each component based on a time series across the orbital cycle.
    However, the procedure necessitates a \vr\ optimisation during the process, enabling us to extract a precise \vr\ for each component in every observation.
    The \vr\ values obtained through this method are presented in Fig.~\ref{fig:vradLSD} and summarised in Appendix~\ref{ap:vr}.

    \begin{figure}
        \centering
        \includegraphics[width=0.99\linewidth]{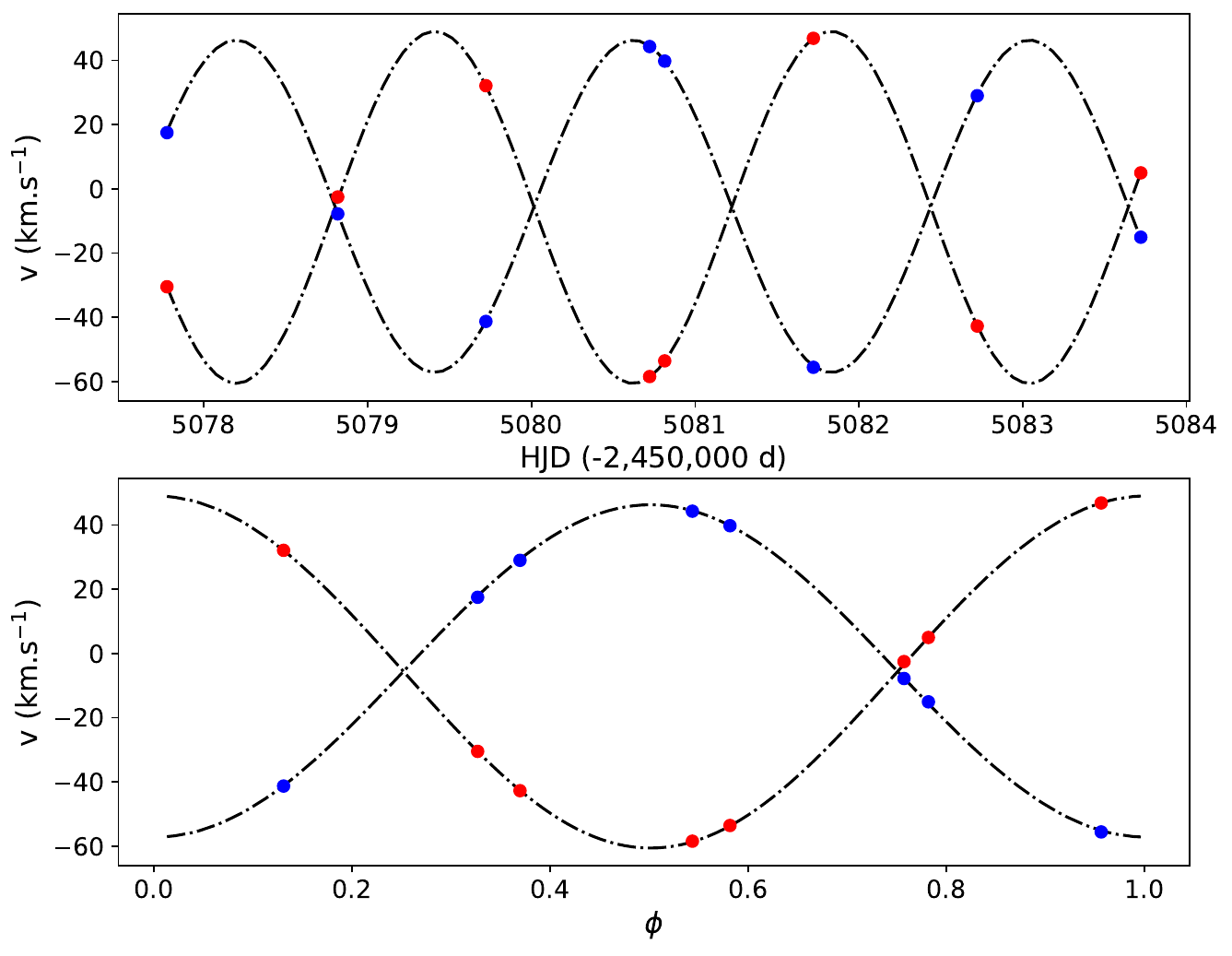}
        \caption{Radial velocity curves of the primary \textit{(blue)} and secondary \textit{(red)} components of V4046 Sgr. The top panel represents the velocity as function of the HJD, while the bottom panel shows the curve folded in phase using the ephemeris of Eq.~\ref{eq:ephemeris}. The dash-dotted curve shows the orbital solution using the parameters of Table~\ref{tab:orbsol}. The error bars are included in this plot but are smaller than the markers' size.}
        \label{fig:vradLSD}
    \end{figure}

    Finally, we utilised a Levenberg-Marquardt algorithm to simultaneously fit an orbital solution to the individual \vr\ curves and their difference, allowing us to derive the orbital parameters of the system.
    The values obtained are summarised in Table~\ref{tab:orbsol}. The derived orbital period, \porb=2.42246$\pm$0.00055~d, is consistent within 1$\sigma$ of the value reported by \cite{Stempels04}, and within 2$\sigma$ of \cite{Quast00}.
    We will thus use the following ephemeris, to define the orbital phases used in this work:
    \begin{equation}
        {\rm HJD} = 2455076.9827 + 2.42246E,
        \label{eq:ephemeris}
    \end{equation}
    where $E$ is the orbital cycle.
    We stress to the reader that this ephemeris is slightly different from \cite{Donati11b} ($T_{\rm 0}$=2\,446\,998.335 and \porb=2.4213459~d), implying small differences in the phases used in this work but yielding conjunctions at phases 0.25 and 0.75 as expected.
    
    \begin{table}
    \centering
    \caption{Orbital elements derived from the \vr\ curves.}
    \begin{tabular}{ll}
        \hline
        Parameter & Value  \\
        \hline 
        \porb (d) & 2.42246$\pm$0.00055 \\
        $\gamma$ (\kms) & $-$5.588$\pm$0.067  \\
        $K_{\rm 1}$ (\kms) & 51.7$\pm$11.8  \\
        $K_{\rm 2}$ (\kms) & 54.8$\pm$12.5  \\
        $e$ & $-$0.0032$\pm$0.0024  \\
        $\omega$ ($\deg$) & 179.4$\pm$26.3  \\
        $T_{\rm 0}$ (HJD$-$2\,450\,000 d) & 5076.9827$\pm$1.8680 \\

        \hline
    \end{tabular}
    \label{tab:orbsol}
    \end{table}

    \subsection{Balmer lines}
    \label{subsec:balmer}

\begin{figure*}
        \centering
        \includegraphics[width=0.33\linewidth]{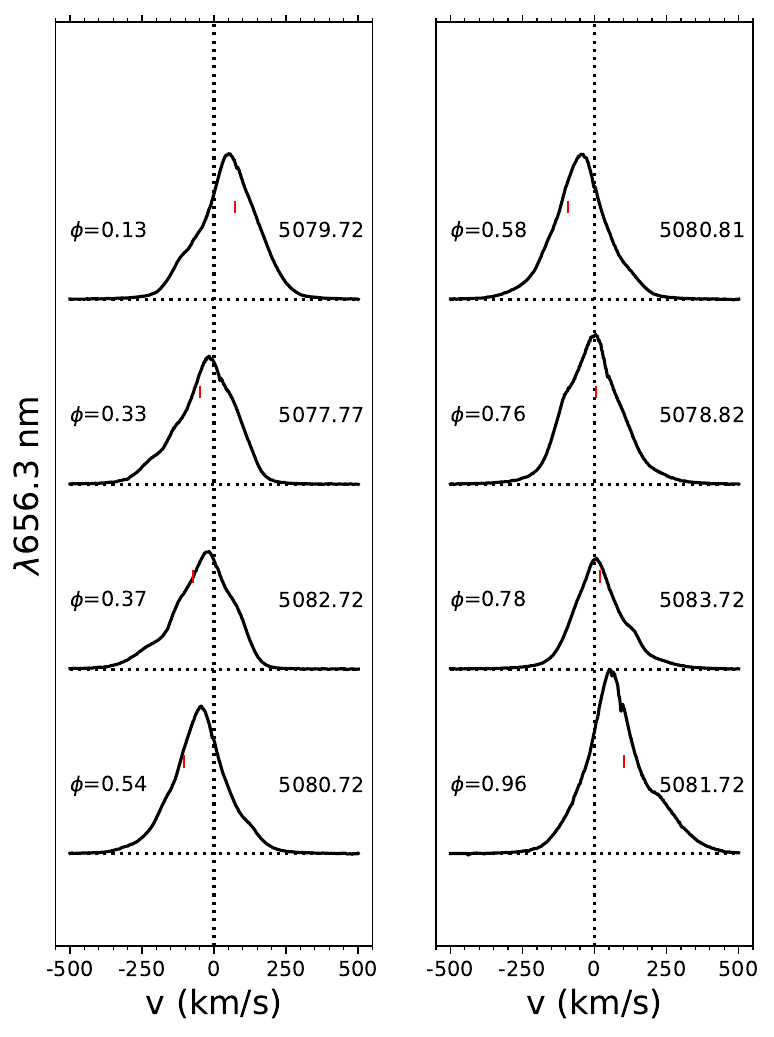}
        \includegraphics[width=0.33\linewidth]{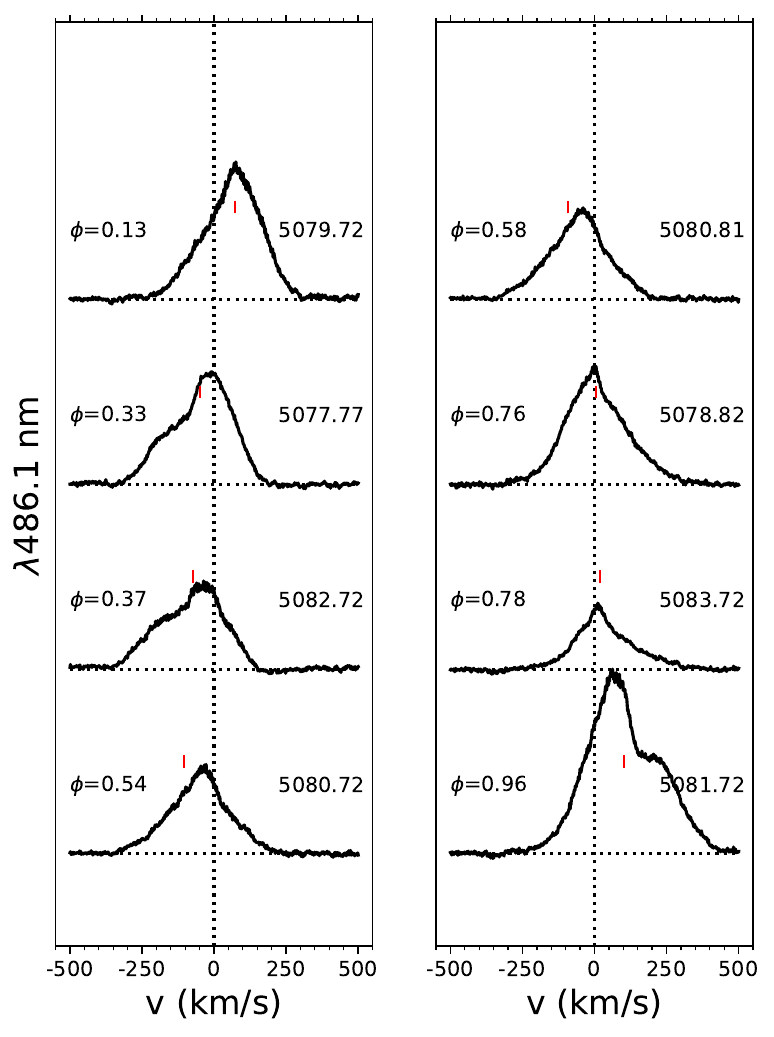}
        \includegraphics[width=0.33\linewidth]{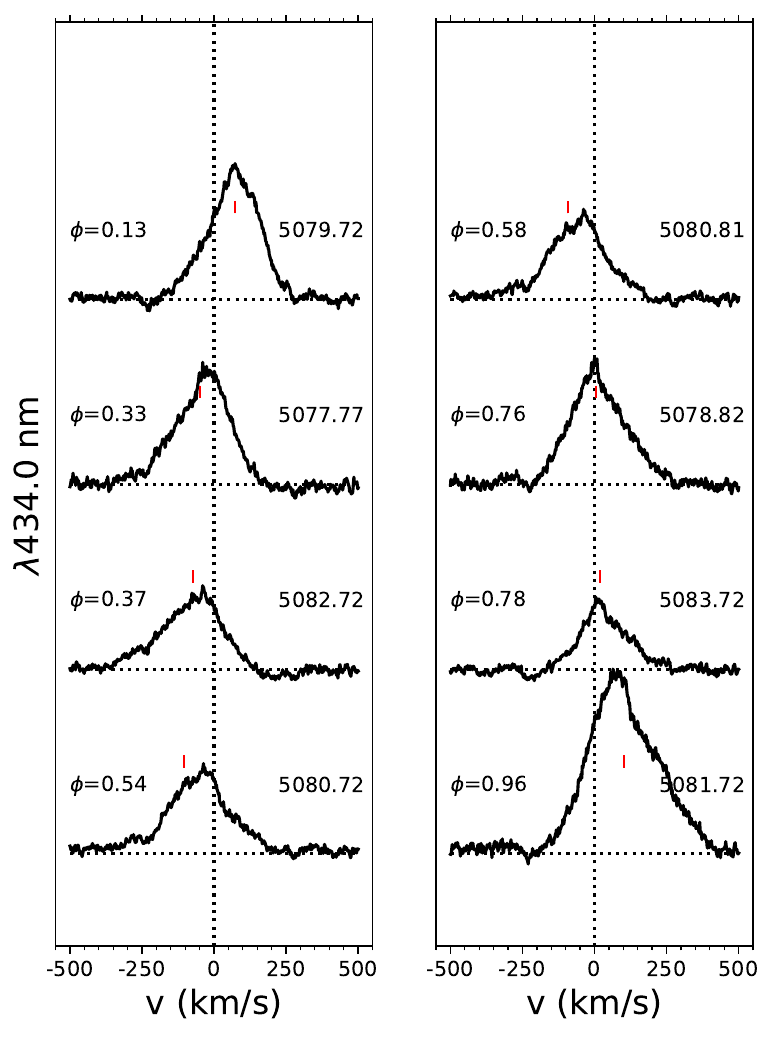}
        \caption{H$\alpha$ (left two pannels), H$\beta$ (middle two panels), and H$\gamma$ (right two panels) residual profiles of V4046~Sgr. These profiles are ordered by orbital phase and corrected from the radial velocity of the primary. The red ticks indicate the velocity of the secondary. The orbital phases and HJDs are indicated on the left and right of each profile, respectively.}
        \label{fig:balmerLines}
    \end{figure*}

    The accretion funnel flows invoked by the magnetospheric accretion process meet the conditions required for the emission of hydrogen lines such as H$\alpha$, H$\beta$, and H$\gamma$ of the Balmer series \citep{Muzerolle01}.
    These lines are therefore suitable for studying the accretion process.
    Figure~\ref{fig:balmerLines} presents the residual profiles of these three lines.

    To compute these profiles, we used a photospheric template with a temperature similar to that of the V4046 Sgr components: V819 Tau, a non-accreting T Tauri star with \vsini=9.5~\kms\ \citep{Donati15}.
    We combined two spectra of V819 Tau, accounting for the luminosity ratio of the V4046~Sgr components \citep[$LR$=1.27,][]{Hahlin22}, to perform the photospheric correction.

    These lines exhibit significant variability and appear to be primarily emitted at the velocity of the secondary component, which may indicate that the secondary is the system’s main accretor.
    In the H$\gamma$ line, one can observe the presence of a redshifted absorption at phases 0.33 and 0.37, extending up to $+$350~\kms.
    This behaviour, known as an inverse P Cygni (IPC) profile, indicates infalling material passing through the line of sight which is characteristic of the magnetospheric accretion process.

    We computed the 2D periodograms of the three lines, consisting of a Generalised Lomb-Scargle periodogram (GLS) calculated in each velocity channel, enabling us to study the periodic modulations within the lines.
    In the primary's velocity frame, the lines behave similarly, exhibiting a periodic variability of the red wing consistent with the orbital period, albeit with a relatively high false alarm probability \citep[FAP=10$^{-1}$, computed from the prescription of][]{Baluev08}.
    Additionally, a signal over the blue wing appears slightly higher in frequency than, but still consistent with, the orbital period (FAP=10$^{-2}$).
    These two signals are consistent with an orbital modulation induced by the emission from the secondary.

    Interestingly, the 2D periodograms also reveal a signal around $f$=0.3~d$^{-1}$ at the centre of the line.
    While the FAP associated with this period is high for H$\alpha$ and H$\beta$ ($>$10$^{-1}$), it is significantly lower for H$\gamma$ (10$^{-2}$).

    In the secondary's velocity frame, only the H$\alpha$ blue wing exhibits a significant periodicity detection, consistent with the orbital period and with a FAP of 10$^{-1}$.
    At the line centre, the signal with the lowest FAP occurs at a frequency of approximately 0.95 d$^{-1}$, which corresponds to the time series sampling.

    \begin{figure*}
        \centering
        \includegraphics[width=0.33\linewidth]{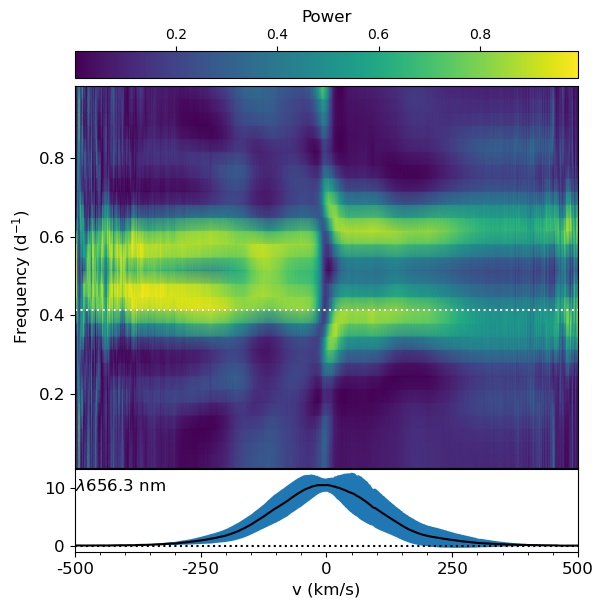}
        \includegraphics[width=0.33\linewidth]{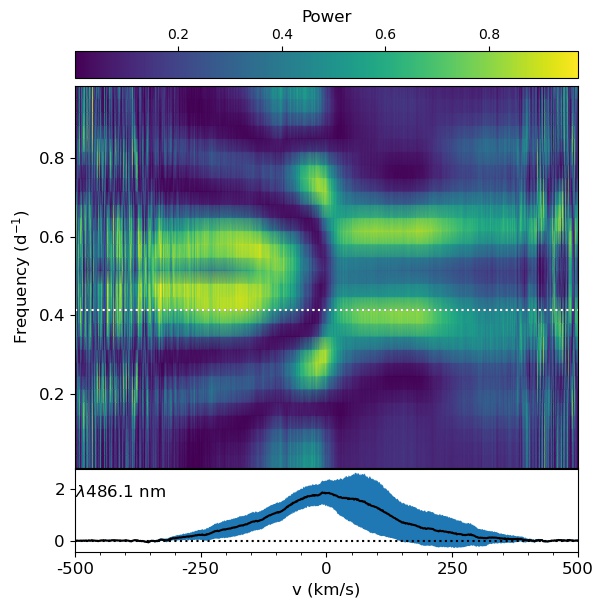}
        \includegraphics[width=0.33\linewidth]{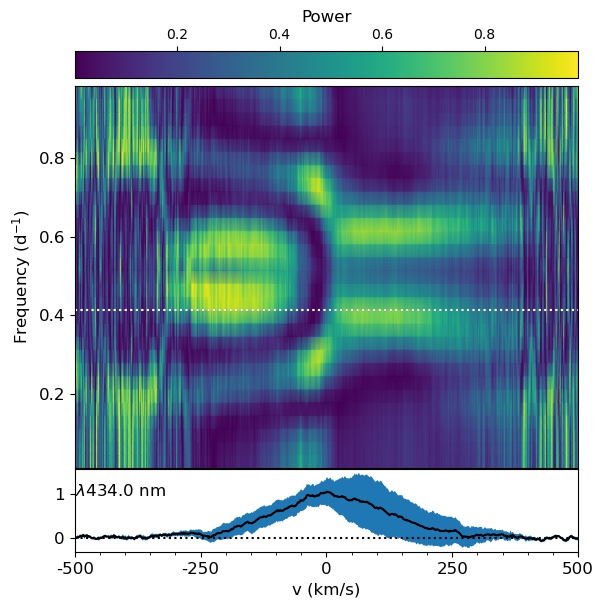}
        \includegraphics[width=0.33\linewidth]{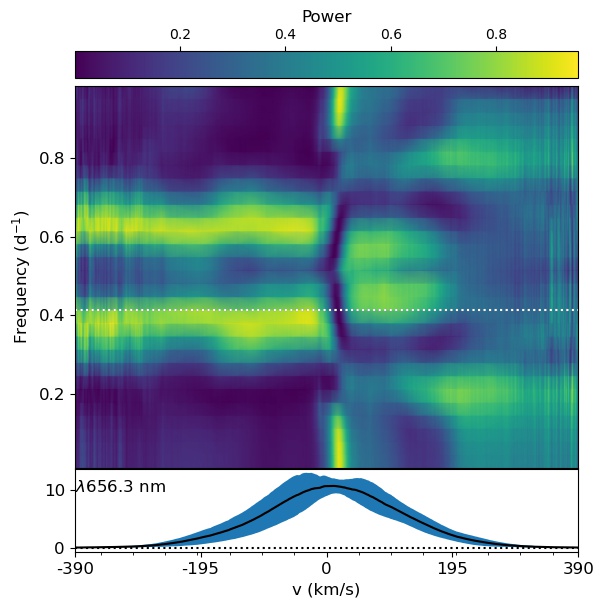}
        \includegraphics[width=0.33\linewidth]{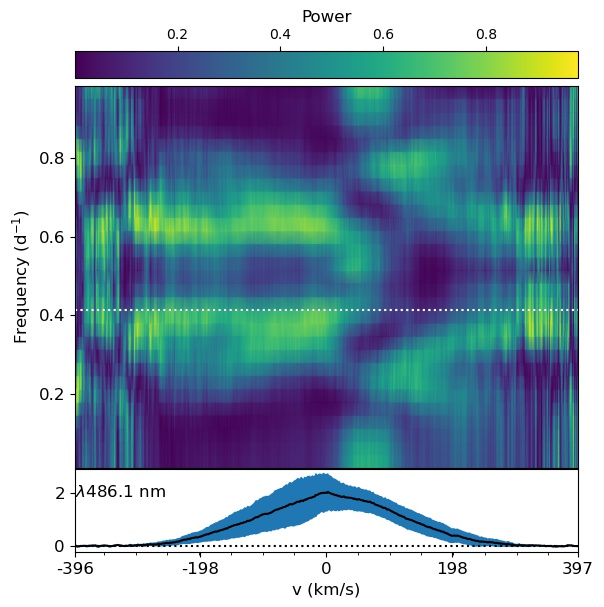}
        \includegraphics[width=0.33\linewidth]{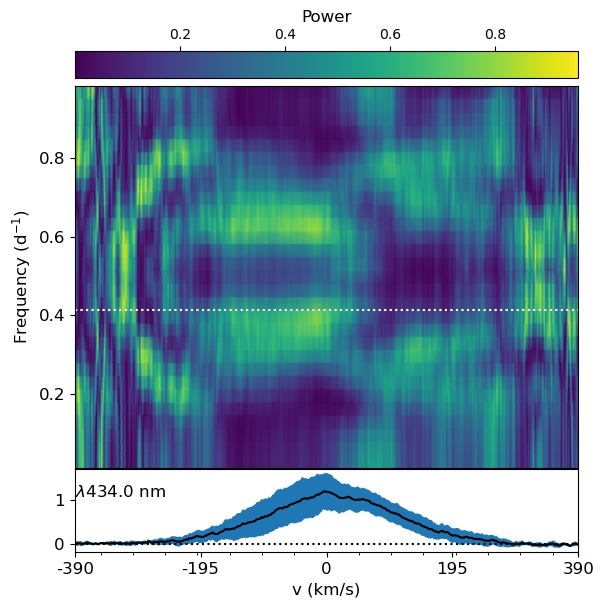}
        \caption{2D periodograms of H$\alpha$ \textit{(left)}, H$\beta$ \textit{(middle)}, and H$\gamma$ \textit{(right)} residual lines in the primary's \textit{(top row)} and secondary's \textit{(bottom row)}  velocity frame. The colour scales the power of the periodogram and the white dotted line highlight the orbital period. The mean profile and it variance are shown at the bottom of each plot in black and blue, respectively. The mirroring effect is due to the 1-day aliasing, a spectral leakage of the Fourrier transform reproducing, with the real signal, the observation sampling. The signal and its alias are disentangled thanks to the FAP.}
        \label{fig:P2DbalmerLines}
    \end{figure*}

    To identify the different variabilities within these lines, we focused on H$\gamma$, which exhibits multiple periodicities, to perform a cross-correlation matrix analysis.
    This tool allows us to identify the correlation between variability regions within a line by calculating a linear correlation coefficient (here, a Pearson coefficient) in each velocity channel of two lines (or twice the same line for an autocorrelation matrix).
    A correlation (near 1) indicates variability dominated by a single physical process, while an anti-correlation (near $-$1) may also indicate two correlated processes.

\begin{figure}
        \centering
        \includegraphics[width=0.49\linewidth]{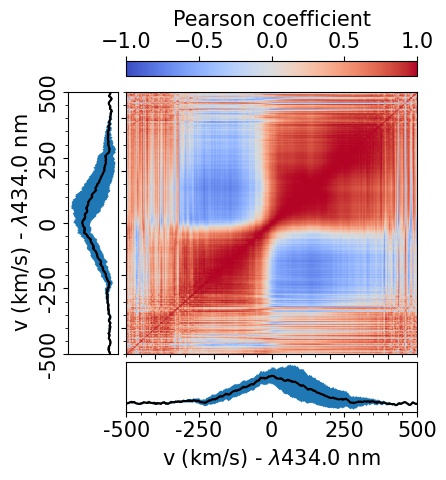}
        \includegraphics[width=0.49\linewidth]{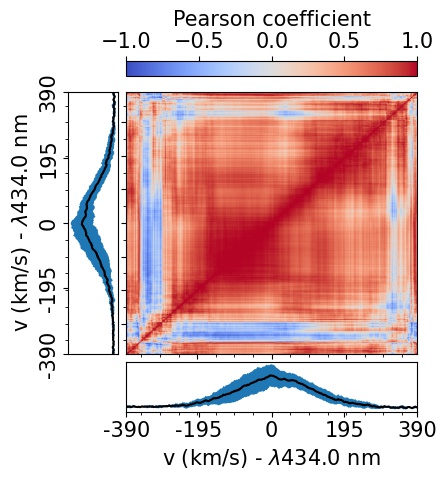}
        \caption{H$\gamma$ residual line auto-correlation matrix in the velocity frame of the primary \textit{(left)} and secondary \textit{(right)}. The colour-code scales the correlation coefficient. Red represents a strong correlation, and blue shows a strong anti-correlation. The plots along the x and y axes represent the corresponding mean line profile (black) and its variance (blue).}
        \label{fig:CMhgamma}
    \end{figure}

    The autocorrelation matrix of H$\gamma$ is shown in Fig.~\ref{fig:CMhgamma}.
    When the line is set in the velocity frame of the primary, the correlation matrix exhibits three main correlated regions (with $r > 0.9$).
    Two of these regions are symmetric around the line centre (between $-$250 and $-$50~\kms, and between $+$30 and $+$200~\kms) and are slightly anti-correlated with each other ($r \sim -0.7$).
    This behaviour typically traces the motion of the emission from the secondary.
    The third correlated region lies between $+$200 and $+$350~\kms, a velocity range consistent with the IPC profile observed at phases 0.33 and 0.37.

    When the line is set in the velocity frame of the secondary, the autocorrelation matrix appears entirely different, as expected if the emission from the secondary differs from that of the primary.
    The three correlated regions are now located between $-$150 and 0~\kms, 0 and $+$80~\kms, and $+$120 and $+$230~\kms.
    
    Finally, we examined the equivalent width (EW) variation. The three lines exhibit similar variations; thus, we present only the results for H$\gamma$ in Fig.~\ref{fig:EWbalmer}. The values for all three lines are summarised in Appendix \ref{ap:ew}.

    Interestingly, the EWs are not modulated by the orbital period and present a clear extremum at phase 0.9. 
    A GLS periodogram identified a period of 3.29~d ($f$=0.304~d$^{-1}$), though with a very high FAP of approximately 0.6, preventing a definitive detection.
    However, this period is consistent with the signal observed at the line centre (see Fig.~\ref{fig:P2DbalmerLines}).
    
    \begin{figure}
        \centering
        \includegraphics[width=.99\linewidth]{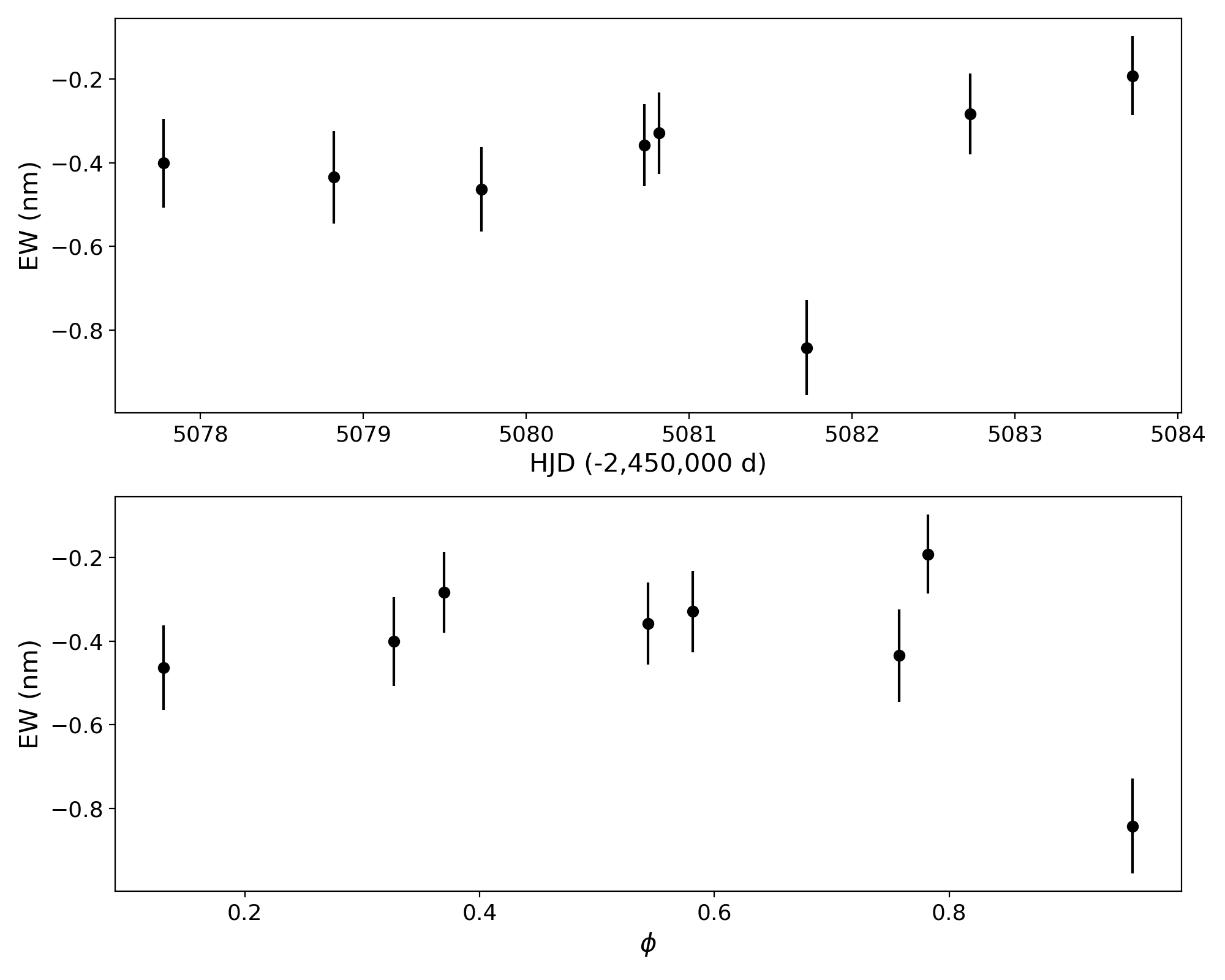}
        \caption{EW of the H$\gamma$ residual line profiles ordered by HJD \textit{(top)} and folded in phase \textit{(bottom)}.}
        \label{fig:EWbalmer}
    \end{figure}

    \subsection{\ion{He}{I} D3}
    \label{subsec:heid3}
    
\begin{figure}
        \centering
        \includegraphics[width=0.99\linewidth]{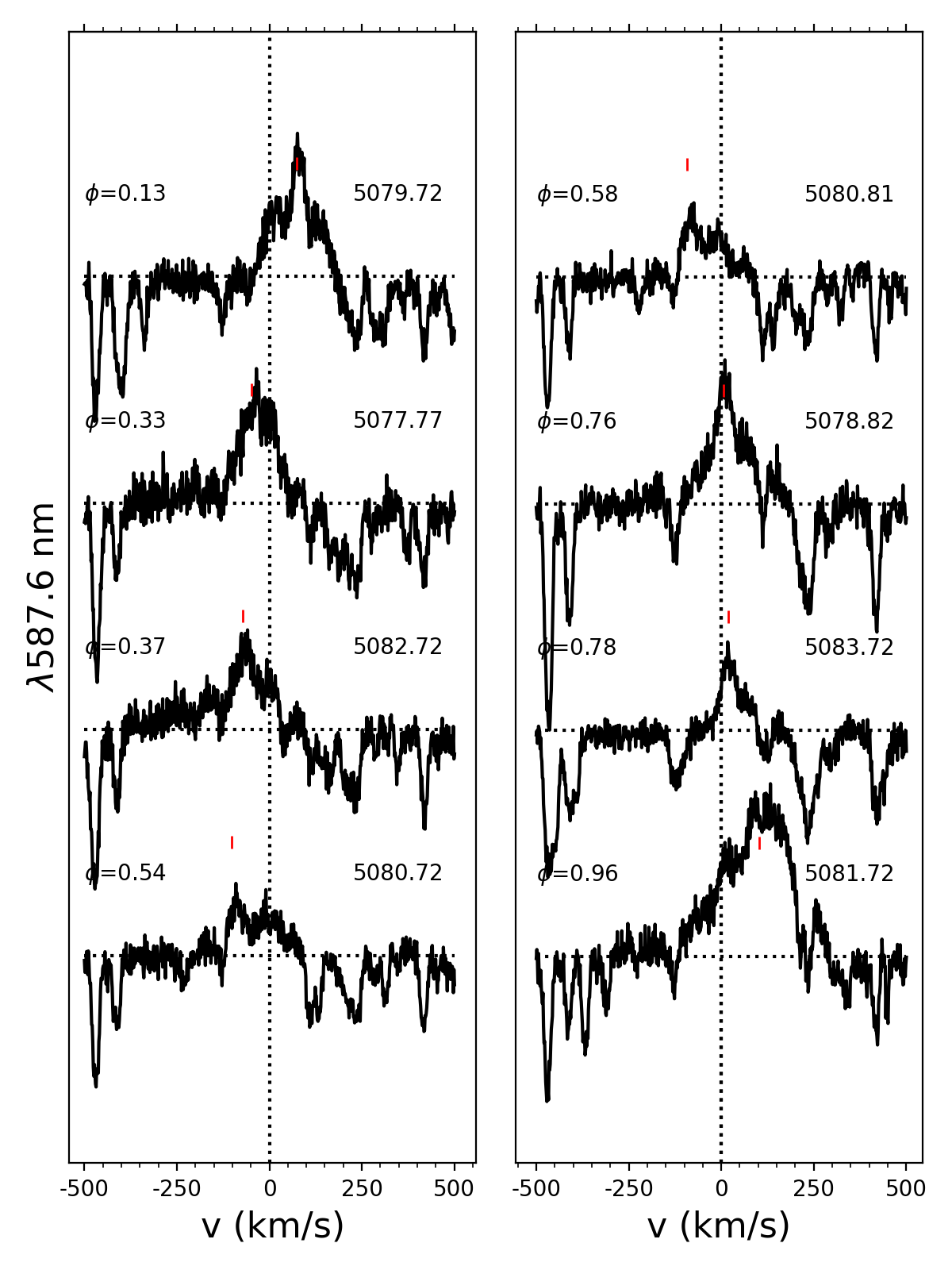}
        \caption{Same as Fig.~\ref{fig:balmerLines} for the \ion{He}{I}~D3 line.}
        \label{fig:heid3}
    \end{figure}

    The narrow component (NC) of the \ion{He}{I} D3 line (587.6nm) is traditionally employed to investigate the magnetospheric accretion process \citep[e.g.,][]{Bouvier20b, Bouvier23, Nowacki23, Pouilly24c}.
    Its formation requires the hot and dense conditions found exclusively in the post-shock region of the accretion shock at the stellar surface \citep{Beristain01}.
    The \ion{He}{I} D3 lines of V4046 Sgr are shown in Fig.~\ref{fig:heid3}.
    Even though these lines suffer from low S/N and strong variability, the presence of an NC is evident at nearly all phases at the secondary's velocity.
    This suggests that the secondary appears to be the main accretor in the system, as also indicated by the Balmer lines (see Fig.~\ref{fig:balmerLines}).
    
\begin{figure}
        \centering
        \includegraphics[width=0.49\linewidth]{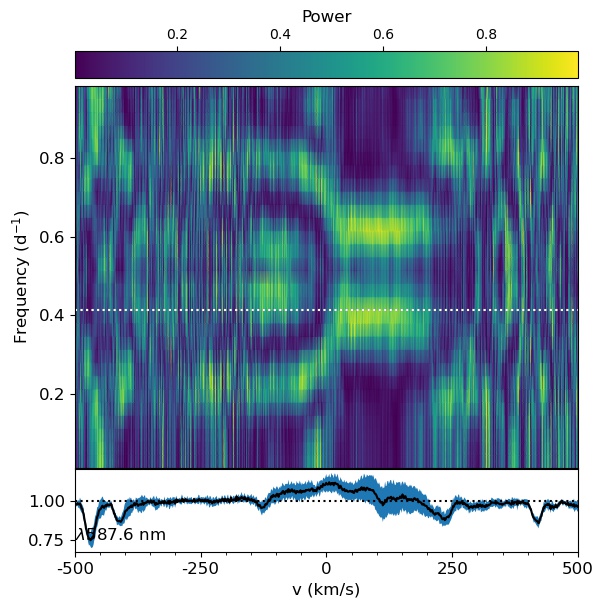}
        \includegraphics[width=0.49\linewidth]{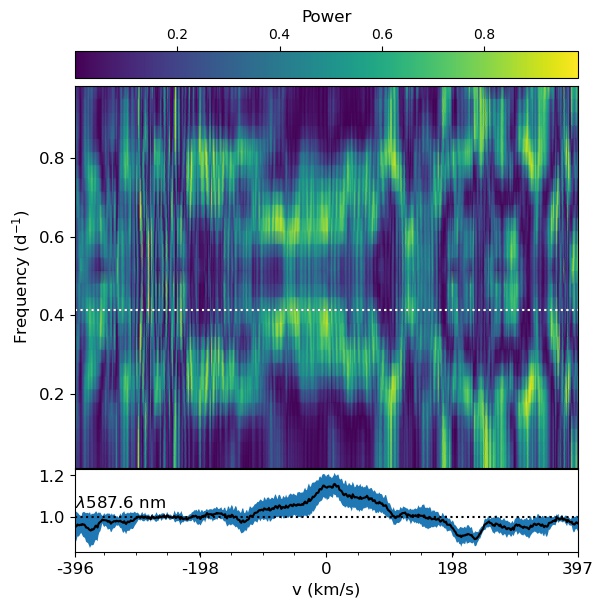}
        \includegraphics[width=0.49\linewidth]{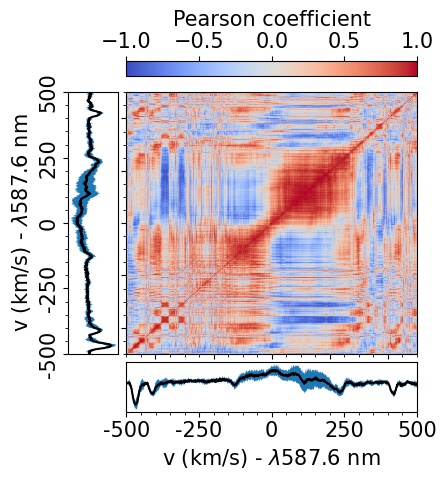}
        \includegraphics[width=0.49\linewidth]{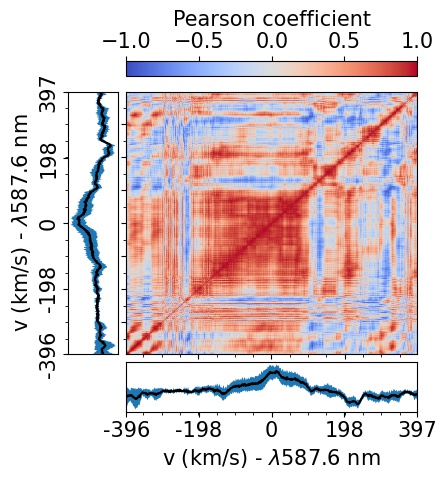}
        \caption{Same as Fig.~\ref{fig:P2DbalmerLines} \textit{(top row)} and Fig.~\ref{fig:CMhgamma} \textit{(bottom row)} for the \ion{He}{I}~D3 line in the primary \textit{(left)} and secondary \textit{(right)} velocity frame.}
        \label{fig:P2DCMheid3}
    \end{figure}

    The periodogram analysis of the line is presented in Fig.~\ref{fig:P2DCMheid3}.
    In the primary's velocity reference frame, the red wing of the line is modulated at a frequency slightly lower than, but consistent with, the orbital period, with a false FAP reaching 10$^{-3}$.
    At the line centre, where the NC of the primary is located, the frequency of modulation appears around $f$ = 0.3~d$^{-1}$ (FAP = 10$^{-1}$), consistent with the signal observed in the Balmer line analysis (see Sect.~\ref{subsec:balmer}).
    In the secondary's velocity reference frame, the signal consistent with the orbital period previously observed in the red wing is shifted towards the line centre (FAP = 10$^{-2}$), and no signal around $f$ = 0.3~d$^{-1}$ with a significantly low FAP is detected.

    Figure~\ref{fig:P2DCMheid3} also presents the auto-correlation matrices of the line in both velocity reference frames.
    As with the Balmer lines, the matrix in the primary's velocity reference frame shows a typical SB2 modulation behaviour, with well-correlated blue and red wings that are anti-correlated with each other.
    In the secondary's velocity reference frame, the matrix exhibits three substructures: between $-$140 and $-$60~\kms, $-$60 and 0~\kms, and $+$20 and $+$75~\kms.
    The negative and positive regions are likely associated with the primary emission, while the region around the line centre corresponds to the NC of the secondary.

    \begin{figure}
        \centering
        \includegraphics[width=0.49\linewidth]{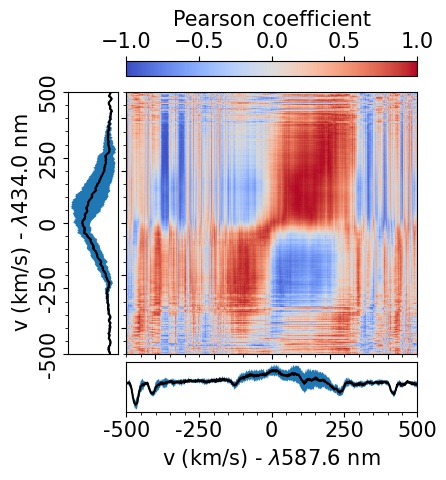}
        \includegraphics[width=0.49\linewidth]{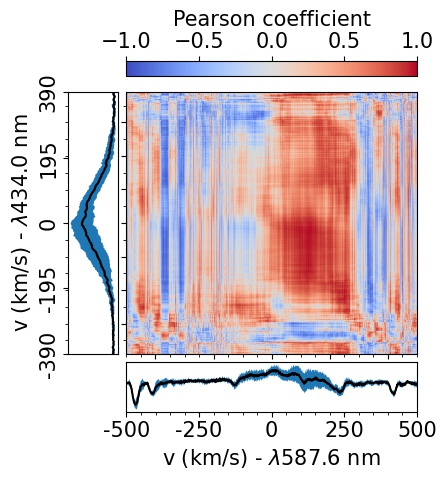}
        \includegraphics[width=0.49\linewidth]{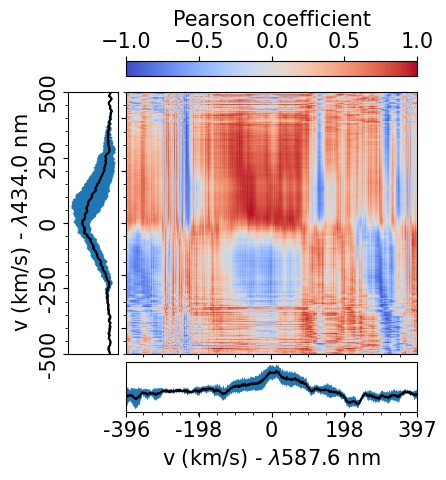}
        \includegraphics[width=0.49\linewidth]{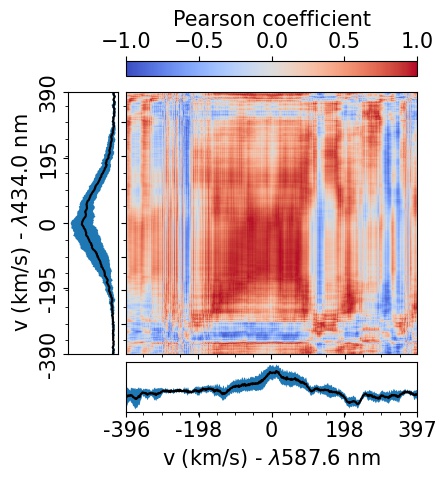}
        \caption{Correlations matrices H$\gamma$ vs. \ion{He}{I}~D3. \textit{Left column}: H$\gamma$ is set to the primary's velocity frame. \textit{Right column}: H$\gamma$ is set to the secondary's velocity frame. \textit{Top row}: \ion{He}{I}~D3 is set to the primary's velocity frame. \textit{Bottom row}: \ion{He}{I}~D3 is set to the secondary's velocity frame.}
        \label{fig:CMheid3_hgamma}
    \end{figure}

    Then we cross-correlated the \ion{He}{i} D3 with the H$\gamma$.
    When both lines are set to the primary's velocity reference frame, the typical SB2 behaviour is recovered.
    When the two lines are set to the secondary's velocity reference frame, the overall \ion{He}{I} D3 line appears to be correlated with the entire blue region of H$\gamma$.
    A detailed analysis of the highest correlation coefficients ($>$0.95) reveals that the $\sim -$60~\kms\ region of the \ion{He}{I} D3 line is highly correlated with the line centre of H$\gamma$.
    Additionally, two other regions of the \ion{He}{I} D3 line, around $+$20 and $+$80~\kms, show high correlation with the blue wing of H$\gamma$ (between $-$60 and $-$160~\kms).

    When the H$\gamma$ line is set to the primary's velocity reference frame and the \ion{He}{I} D3 line to the secondary's velocity reference frame, the correlation matrix exhibits different behaviour.
    A strong correlation ($>$0.95) is observed between the \ion{He}{I} D3 line (at $\sim -$60~\kms) and the red wing of H$\gamma$ (between $+$30 and $+$120~\kms).
    The anti-correlation between the \ion{He}{I}~D3 line and the blue wing of H$\gamma$ is weaker, with coefficients around $-$0.7.

    In the opposite velocity frame configuration, only a redshifted region of the \ion{He}{I}~D3 line (around $+$130\kms) exhibits a strong correlation with the blue wing of H$\gamma$.

    Although these matrices may seem disorderly, they suggest accretion signatures in H$\gamma$ and \ion{He}{I}~D3 from both components. 
    Some accretion occurs along the line of sight at the same time, producing the observed correlations between the line centres and the blue- and red-shifted regions.

    \subsection{\ion{Ca}{II} infrared triplet}
    \label{subsec:caIIirt}

    Given the limited information we could extract from the NC of the \ion{He}{I} line, we extended our investigation by examining the \ion{Ca}{II} infrared triplet (IRT), located at approximately 849.8, 854.2, and 866.2~nm.
    Given the similar shape and variability of the three lines, we first performed an LSD-like averaging of the three lines to produce a profile that we refer to as the \ion{Ca}{II}~IRT line for simplicity.

\begin{figure}
        \centering
        \includegraphics[width=0.99\linewidth]{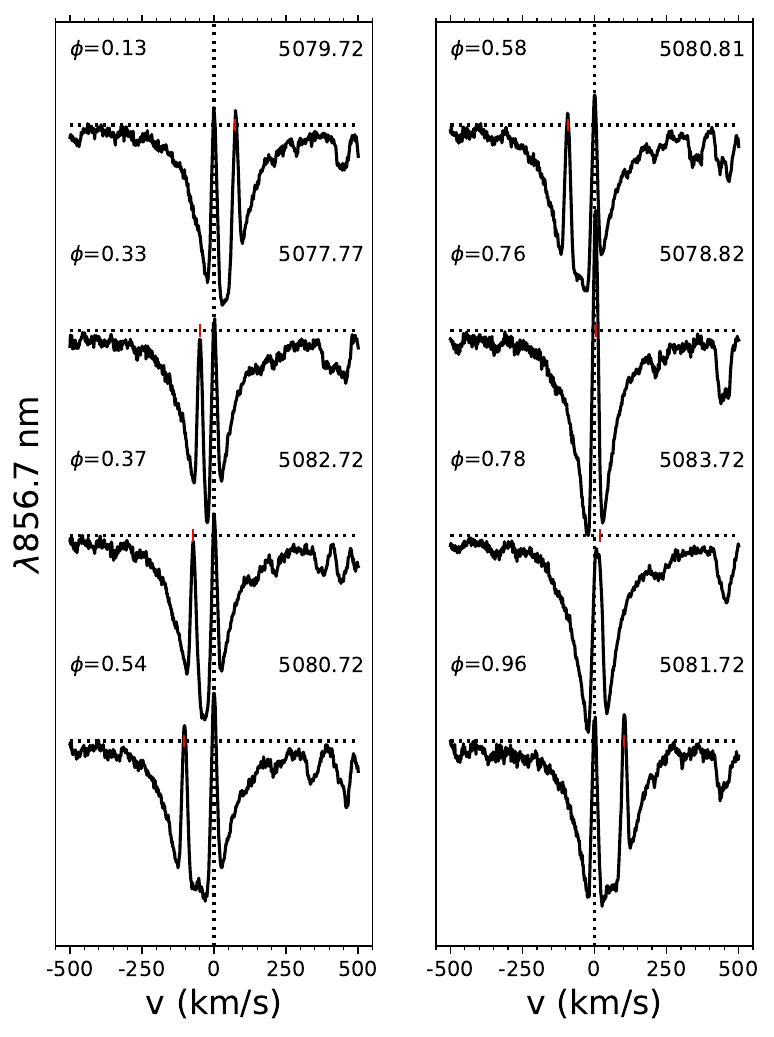}
        \caption{Same as Fig.~\ref{fig:balmerLines} for the \ion{Ca}{II}~IRT line profiles.}
        \label{fig:caiiirt}
    \end{figure}

    These lines are shown in Fig.~\ref{fig:caiiirt} and consist of the photospheric absorption of the two components, as well as two NCs in emission, originating close to the stellar surface.
    Studying the NCs of the \ion{Ca}{II}~IRT allows us to trace the accretion process of each component.
    Surprisingly, a 2D periodogram analysis (not shown here) did not reveal any periodicity in the NCs, whether in the primary's or secondary's velocity reference frame.
    
    To isolate the NCs, we first performed the same double Lorentzian profile fit as \cite{Donati11b} on the photospheric absorption, combined with a double Gaussian profile for the NCs.
    We then applied the same iterative method as for the LSD profiles (see Sect.~\ref{subsec:LSDVr}) to disentangle the two components and produce a mean NC for both the A and B components.
    Finally, we corrected the profiles using the mean NC of the B component to obtain the individual NCs of the A component, and vice versa.
    
    We extracted the \vr\ of the two NCs from the disentangling procedure and fitted the corresponding curves using the method described in \cite{Pouilly21} to constrain the emitting regions of the NCs.
    The results are summarised in Table~\ref{tab:fitVrNC}.
    For the emitting region of the NC of the primary, our results are consistent within 1$\sigma$ with those of \cite{Donati11b}, indicating a location close to the rotation pole and facing the observer at $\phi$=0.66 (the authors found 0.1, but used a different ephemeris, which corresponds to 0.67 in our work).
    However, our results for the NC of the secondary deviate from those of \cite{Donati11b}.
    While we confirm the polar location of the emitting region, it faces the observer at $\phi$=0.01, whereas the latter authors reported $\phi$=0.82 (using our ephemeris), indicating consistency at only 2$\sigma$.
    The large uncertainties on the colatitudes obtained for both components reflect an extended emitting region, also recovered by \cite{Donati11b}.
    
    \begin{table}
        \centering
        \caption{Results of the fit of the \ion{Ca}{ii} IRT NCs velocities}
        \begin{tabular}{lll}
            \hline
            Parameter & NC(A) & NC(B)  \\
            \hline
            $V_{\rm flow}$ (\kms) & 0.67$^{+0.42}_{-0.22}$ & 0.42$^{+0.34}_{-0.51}$ \\
            $V_{\rm eq}$ (\kms) & 20.4 $\pm$9.7 & 18.8$\pm$9.0 \\
            $\phi_{\rm s}$ & 0.66$^{+0.08}_{-0.05}$ & 0.01$^{+0.15}_{-0.09}$ \\
            $\theta$ ($^\circ$) & 5.6$^{+69.3}_{-3.0}$ & 5.0$^{+70.0}_{-3.3}$ \\
            $i$ ($^\circ$) & 30.1$\pm$15.0 & 30.0$\pm$15.0 \\
            \hline
        \end{tabular}
        \tablefoot{$V_{\rm flow}$ is the velocity of the material in the emitting region, $V_{\rm eq}$ is the equatorial velocity, $\phi_{\rm s}$ is the phase where the emitting region faces the observer, $\theta$ is the colatitude of the emitting region, and $i$ is the inclination of the rotation axis.}
        \label{tab:fitVrNC}
    \end{table}

    \begin{figure}
        \centering
        \includegraphics[width=0.99\linewidth]{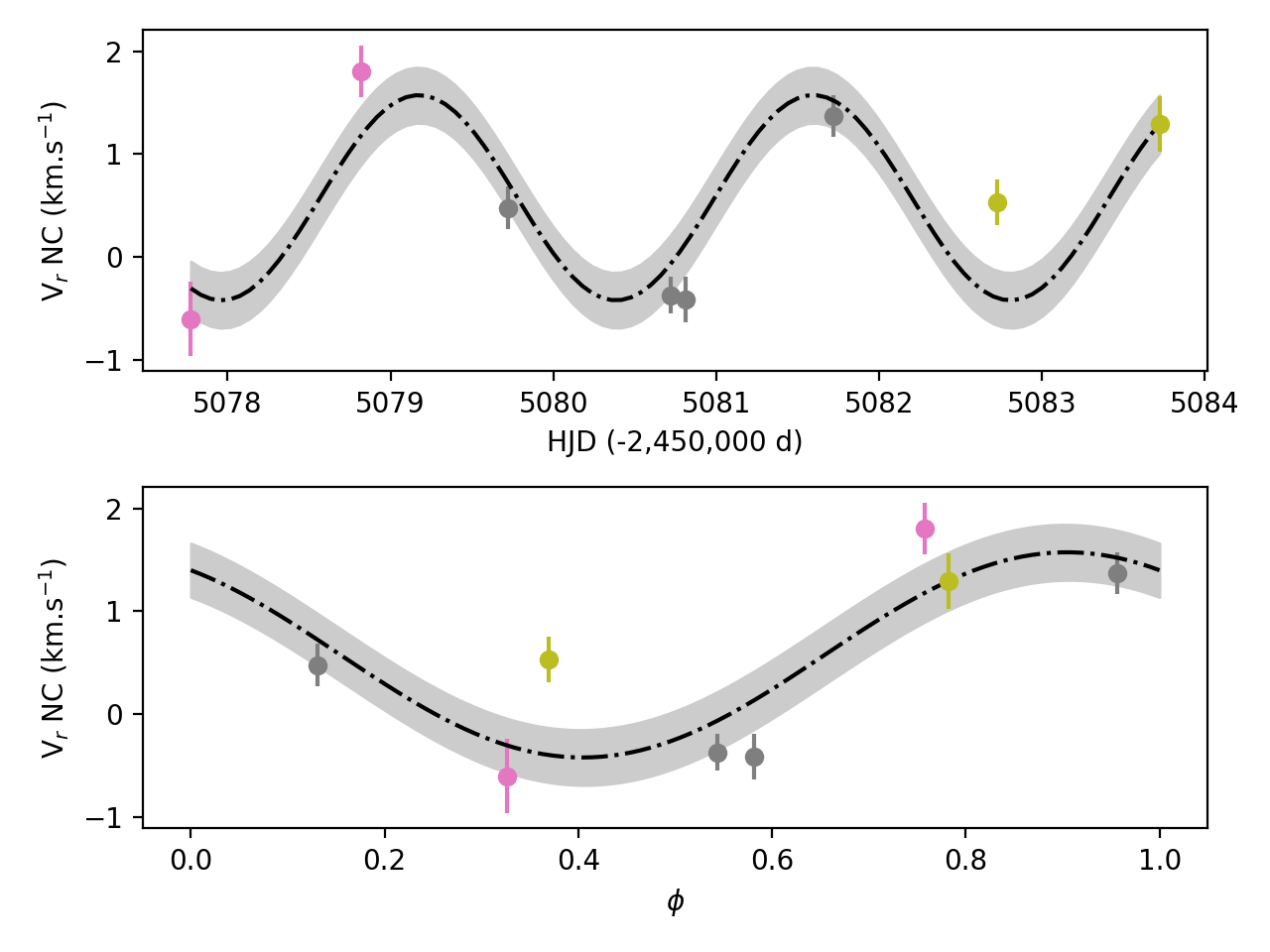}
        \includegraphics[width=0.99\linewidth]{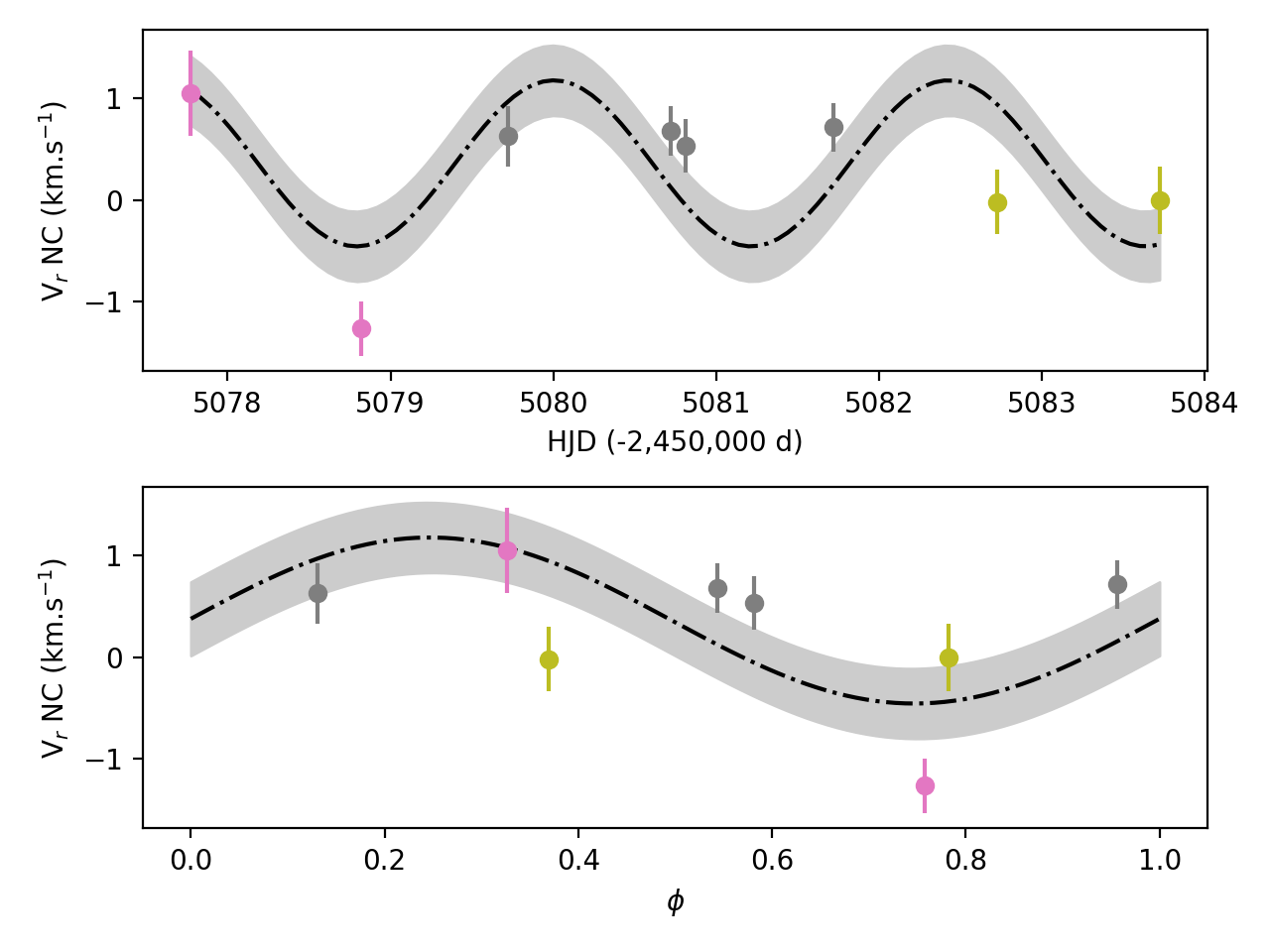}
        \caption{Radial velocity fit of the \ion{Ca}{II}~IRT NC A \textit{(two top panels)} and B \textit{(two bottom panels)} corrected from the Doppler velocity modulation induced by the orbital motion derived in Sect.~\ref{subsec:LSDVr}. Each colour represents an orbital cycle. The black dash-dotted curve is the curve fitted, its uncertainty on the amplitude is represented in grey.}
        \label{fig:NCAVr}
    \end{figure}

\begin{figure}
        \centering
        \includegraphics[width=0.49\linewidth]{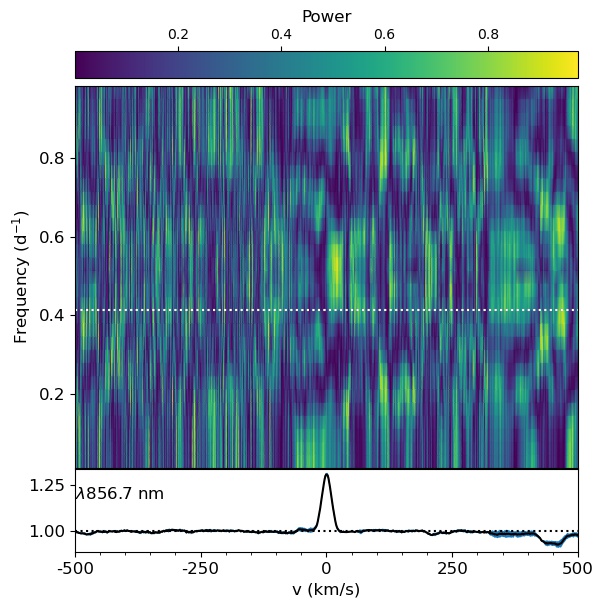}
        \includegraphics[width=0.49\linewidth]{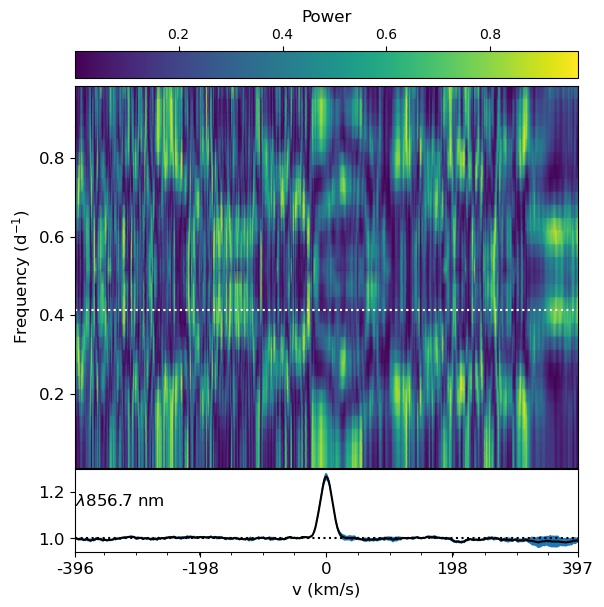}
        \caption{Same as Fig.~\ref{fig:P2DbalmerLines} for the NC of the A \textit{(left)} and B \textit{(right)} components is their respective velocity reference frame.}
        \label{fig:P2DcaNC}
    \end{figure}

    The 2D periodograms of each component's NC in their respective velocity frames are shown in Fig.~\ref{fig:P2DcaNC}.
    The A component shows a signal at $f$=0.5 (FAP=10$^{-2}$).
    The B component appears to be periodic on the same period as the orbital motion, with a fairly low FAP (10$^{-1}$), although the signal seems slightly blue-shifted ($-$25~\kms).

    The aim of disentangling the two NCs was to study the EW of each NC.
    The plots are shown in Fig.~\ref{fig:EWcairtNC}, and the values are summarised in Appendix~\ref{ap:ew}.
    No periodicity was detected in the EWs.
    When examining the EWs of the non-disentangled NCs, two extrema are present at approximately $\phi$=0.3 and 0.8, as expected, since the two components are merged at these phases.
    However, for the EWs of the disentangled NCs, only the extremum at 0.3 remains, indicating that this peak is likely due to an NC feature.
    
    \begin{figure}
        \centering
        \includegraphics[width=0.99\linewidth]{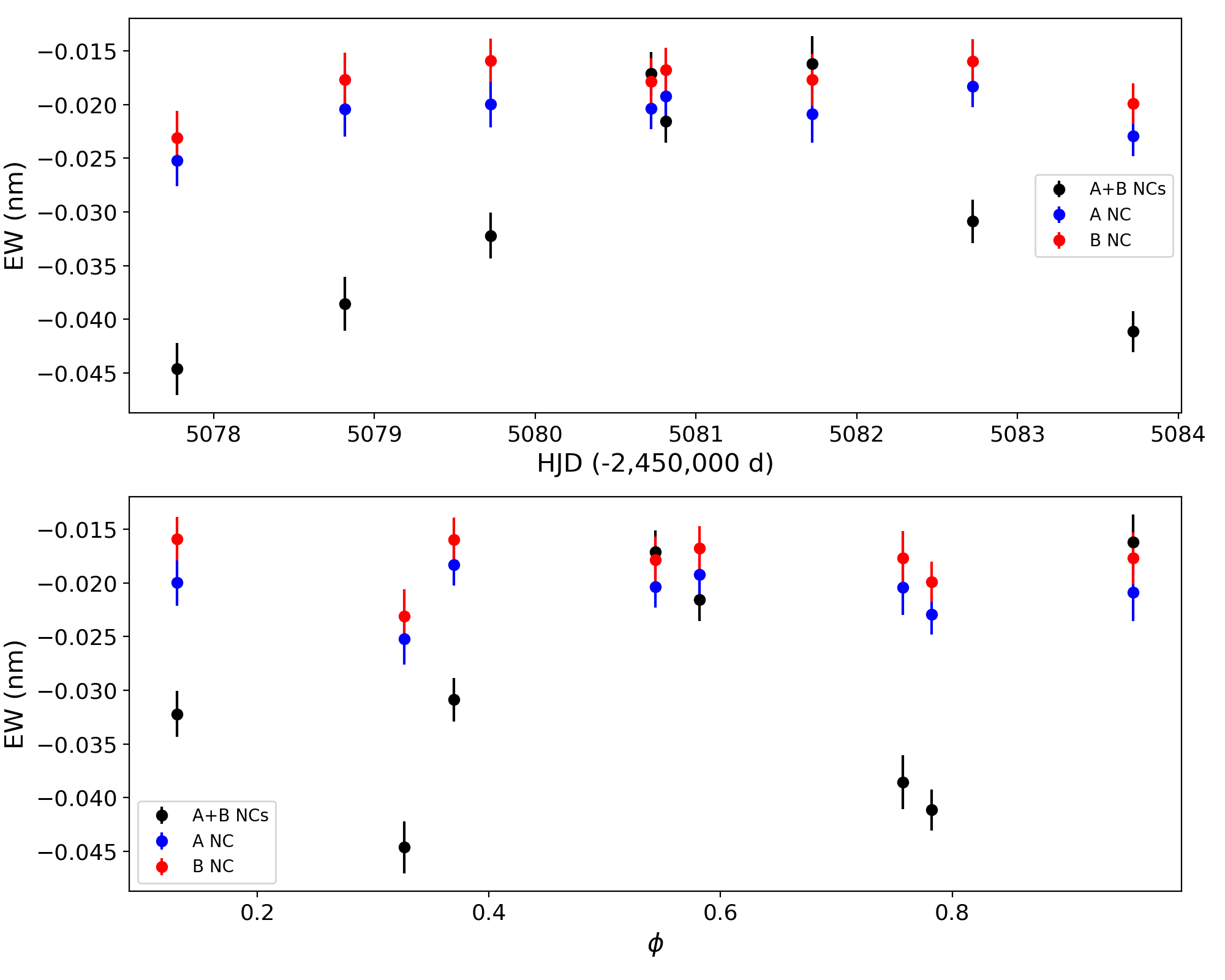}
        \caption{Same as Fig.~\ref{fig:EWbalmer} for the \ion{Ca}{II} IRT NCs. The black dots show the values for the combined profiles, the EWs of the primary's and secondary's NCs are displayed in blue and red, respectively. }
        \label{fig:EWcairtNC}
    \end{figure}

    \subsection{TESS light curves}
 
\begin{figure*}
        \centering
        \includegraphics[width=0.49\linewidth]{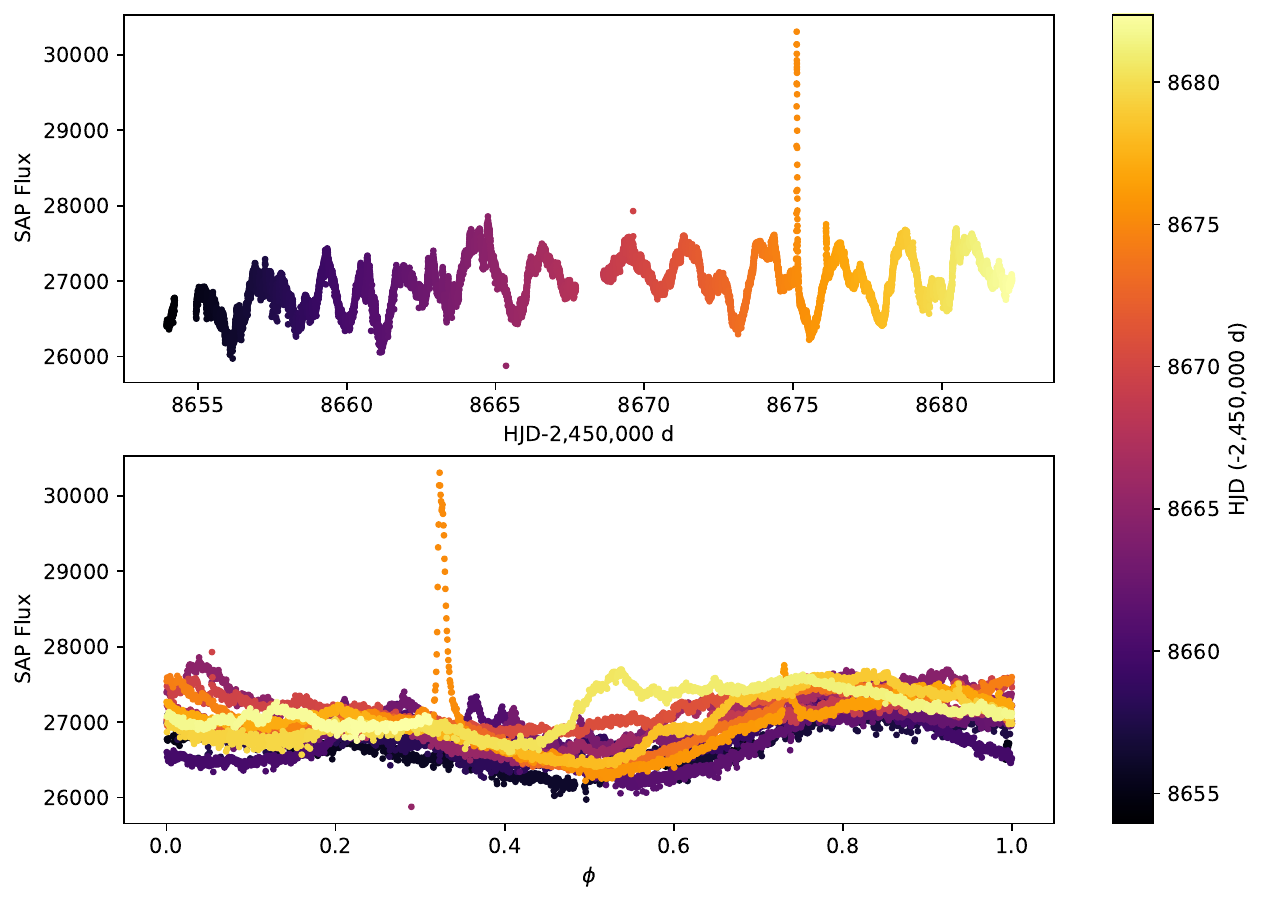}
        \includegraphics[width=0.49\linewidth]{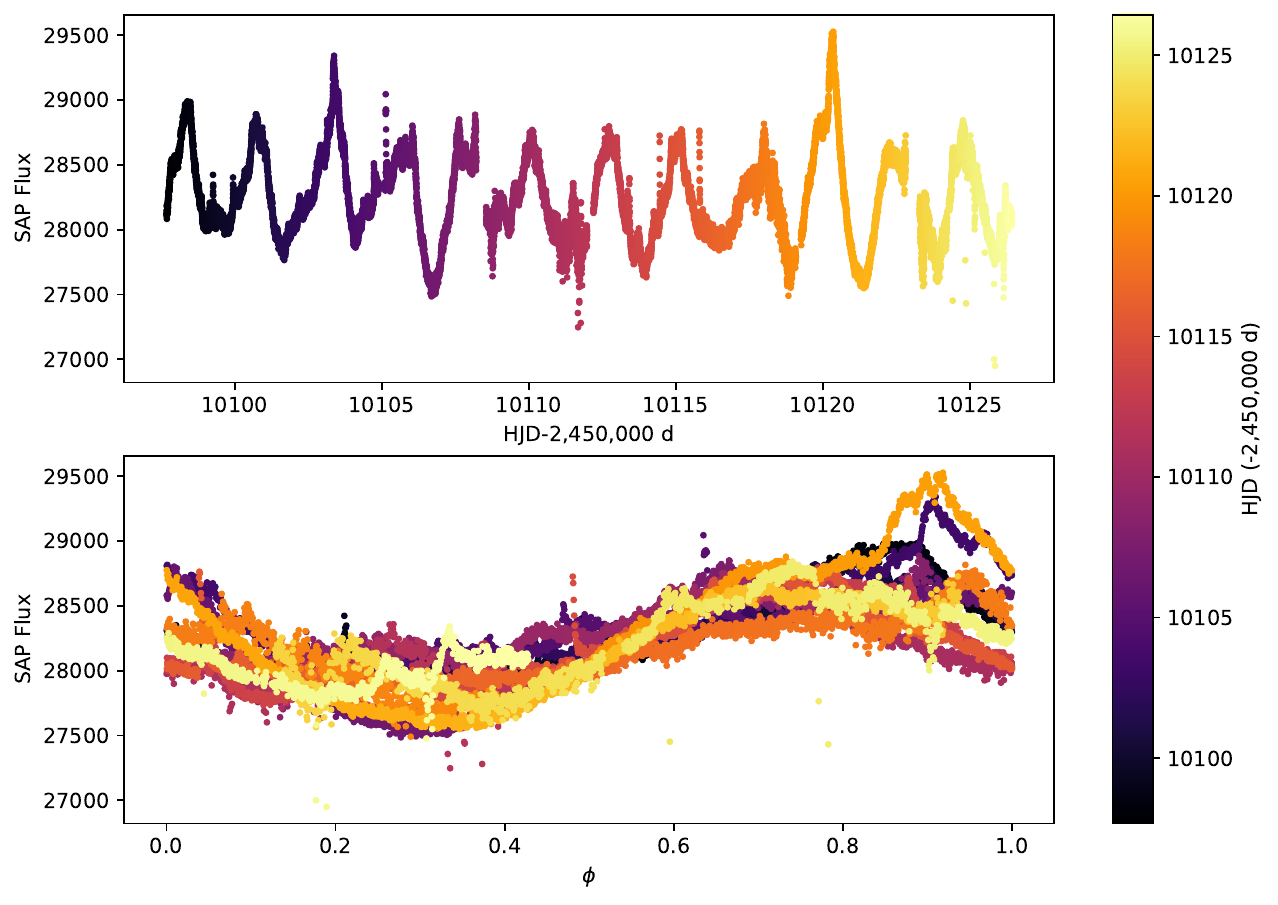}
        \caption{Sector 13 \textit{(left)} and 66 \textit{(right)} TESS light curves \textit{(top panels)}, folded in phase following Eq.~\ref{eq:ephemeris} \textit{(bottom panels)}. The colour scale the HJDs.}
        \label{fig:TESSlc}
    \end{figure*}

    V4046 Sgr was observed by TESS twice, in Sector 13 and 66.
    The two curves of PDC SAP flux are shown in Fig.~\ref{fig:TESSlc}.
    We converted the BTJD to HJD by applying a shift of 69.184s \citep{Nelson23}, and we folded them in phase using the ephemeris in Eq.~\ref{eq:ephemeris}.
    The GLS periodograms analysis yield period detections fully consistent with the orbital period.
    In Sector 13, an isolated flare occurs around HJD 2\,458\,675 ($\phi$=0.35), as well as an enhancement at phase 0.55, which reaches the maximum of luminosity modulation at the point when it should be at its minimum.
    However, in Sector 66, the strongest luminosity enhancement occurs around phase 0.9, on two different cycles, and well above the maximum of the modulation.
    Figure~\ref{fig:phaseTESSlc} also displays the Sectors 13 and 66 TESS light curves folded in phase, but the rotation cycles have been split and shifted vertically to improve the readability of each cycle.
    The phases 0.25, 0.5, and 0.75 are emphasised by vertical grey dotted lines, as they represent particular points of the orbit following the ephemeris in Eq.~\ref{eq:ephemeris}.
    Indeed, $\phi$=0.25 is the point when the radial velocity of the primary switches from negative to positive values (with respect to the systemic velocity), meaning that this component is in front of the observer and the secondary is behind, from phases 0.0 to 0.5.
    Phase 0.75 represents the opposite situation, meaning that the secondary is in front of the observer from phases 0.5 to 1.0.
    We stress to the reader that, given the low inclination of the orbit ($\sim$35$^\circ$), no eclipse occurs and both components are visible at all phases; the two conjunctions only indicate which component is the closest to the observer.
    The phased light curve seems to be a superposition of a typical sinusoidal binary modulation (see cycles 1483 and 2082), with punctual luminosity enhancements due to the accreting component during the different orbital cycles.
    However, the minimum of the modulation is slightly offset from phase 0.25, and the maximum from phase 0.75, as expected if one component is less bright than the other.
    Furthermore, the two components have very similar temperatures, and from the literature, the hottest one is the primary, with a K5 spectral type, the secondary being K7 \citep{Quast00}.
    Assuming a blackbody emission, this means that the minimum should occur at phase 0.75 and the maximum at phase 0.25.
    The sinusoidal modulation observed is thus likely due to a spot on one of the components.
    Concerning the superimposed punctual luminosity enhancements: during cycles 1479 and 1480 (Sector 13), the light curve shows multiple luminosity peaks around phase 0.25, meaning that they represent accretion events on the primary during this cycle.
    However, cycles 1486 and 1487 show an overall luminosity increase centred on phase 0.75, reflecting an accretion event on the secondary.
    When the multiple peaks of these events on the primary indicate a discrete accretion pattern, the process on the secondary seems more continuous, distorting the symmetry of the sinusoidal shape of the modulation.
    This behaviour is confirmed in Sector 66, with multiple luminosity peaks on the primary "side" of the phased curve (cycles 2078, 2079), and a more continuous enhancement on the secondary "side" (cycles 2074, 2083).

\section{Discussion}
\label{sec:discussion}
 
    V4046 Sgr was selected for this study due to the intriguing contradiction between its highly ordered configuration—two components with similar masses, ages, and temperatures, in a circularised orbit with synchronised rotation—and its chaotic accretion process as reported in prior studies.
    While the ESPaDOnS spectropolarimetric time series allowed \cite{Donati11b} to derive the magnetic field topology of the system, no detailed description of its accretion pattern was provided, despite the suitability of such a dataset for this purpose \cite[e.g.,][]{Pouilly20, Pouilly21, Pouilly23, Pouilly24, Pouilly24c}.
    This work seeks to address this gap.

    We first derived the LSD profiles of V4046 Sgr to extract the radial velocity modulation of the system and determine its orbital parameters.
    Our results are fully consistent with those of \cite{Stempels04}, although using their ephemeris would result in a 0.069-cycle shift for conjunctions \citep{Donati11b}.
    This discrepancy was corrected using the orbital parameters derived in this work, which we used to define the ephemeris (Eq.~\ref{eq:ephemeris}), despite the larger uncertainties.

    Our analysis confirmed accretion onto both components, as evidenced by emission line peaks at the velocities of the Balmer lines (tracing accretion funnel flows), \ion{Ca}{II} IRT NC, and \ion{He}{I} NC \citep[formed near the accretion shock,][]{Beristain01}.
    However, we identified significantly different accretion patterns between the two components, as revealed by the variability in line profiles when analysed in the reference frame of each star.

    \subsection{Primary component (V4046 Sgr A)}
    
    An IPC profile in H$\gamma$ at phases 0.33 and 0.37, extending up to $+$350~\kms, indicates the presence of an accretion funnel flow.
    This signature is attributed to the primary based on the line region's periodicity consistent with its rotation period (Fig.~\ref{fig:P2DbalmerLines}) and its clear and separated correlation in the autocorrelation matrix when set to the primary’s velocity reference frame (Fig.~\ref{fig:CMhgamma}).
    According to the magnetospheric accretion scheme for cTTSs \citep[see review by][]{Hartmann16}, such a funnel flow should channel material onto the stellar surface near the dipole pole, forming an accretion shock.

    For V4046 Sgr A, this is consistent with the dipole pole's location at $\phi$ = 0.37 \citep[][using our ephemeris]{Donati11b}, in phase with the IPC profile and the minimum \ion{Ca}{II} IRT NC EW.
    Furthermore, TESS light curves (Fig.~\ref{fig:TESSlc}) reveal multiple accretion-induced luminosity peaks near the time of the primary’s conjunction, likely caused by a discrete accretion flow separated into sub-arms, as observed in systems like V807 Tau \citep{Pouilly21}.

    \subsection{Secondary component (V4046 Sgr B)}

    Accretion onto the secondary appears less structured than that onto the primary.
    The \ion{He}{I} D3 NC indicates the presence of an accretion shock at the stellar surface, periodic with the orbital period, and suggests that V4046 Sgr B is the system's main accretor, because it is mostly located at the seconary's velocity.
    However, the H$\gamma$ EW, thus expected to be dominated by the secondary, shows no periodicity consistent with the rotation period, and its autocorrelation matrix does not exhibits the behaviour expected from the magnetospheric accretion scheme.

    None of the diagnostics aligns with the dipole pole of the secondary \citep[$\phi \approx$ 0.6 using our ephemeris,][]{Donati11b}.
    Moreover, TESS light curves show an overall luminosity enhancement at the phases of the secondary’s conjunction, indicating a less concentrated accretion structure compared to the primary.

    \subsection{Dissimilar accretion patterns}

    This work demonstrates different accretion onto the two components of V4046 Sgr.
    While surprising given their similarities, this difference is consistent with their distinct magnetic topologies.
    V4046 Sgr's evolutionary stage, slightly more advanced than typical cTTSs, has resulted in the development of radiative cores and complex magnetic topologies \citep{Donati11b}.

    Both stars exhibit weak large-scale magnetic fields, with significant toroidal, quadrupolar, and octupolar components, deviating from the strong poloidal and dipole fields typically observed in cTTSs.
    Despite this complexity, magnetospheric accretion can still occur, as seen in HQ Tau \citep{Pouilly20}.
    However, the obliquities of the magnetic dipole axes (60$^\circ$ for V4046 Sgr A and roughly 90$^\circ$ for V4046 Sgr B) are unusually high compared to the 5-to-20$^\circ$ range observed in cTTSs with stable accretion \citep{Mcginnis15}.
    Dipole magnetic obliquity is a key parameter affecting directly the magnetospheric radius, and so the limit between a "stable" regime, where an accretion funnel flow connect the disc to the stellar dipole pole, and and unstable one, characterised by accretion tongues penetrating the magnetosphere and connecting the disc to the stellar equator \citep{Romanova15}.
    
    \subsection{Magnetospheric radii and accretion regimes}

    Using the values from \cite{Donati11b}, the magnetospheric radii are 0.82 ± 0.67 R$_\star$ and 0.09 ± 0.08 R$_\star$ for the primary and secondary (using 89$^\circ$ instead of 90$^\circ$ that brings the value down to 0), respectively.
    For V4046 Sgr B, the nearly perpendicular dipole field likely prevents stable magnetospheric accretion, as its magnetospheric radius does not reach the stellar surface in the disc's plan.

    The primary's magnetospheric radius cannot directly truncate the circumbinary disc \citep[located at $\sim$10 R$\star$,][]{Stempels04}.
    However, the gas bulks detected in the Roche lobes, near the co-linear Lagrangian points, at $\sim$6.93 R$_\odot$ from the system's centre of mass (meaning at $\sim$1.15 R$_\star$ from the stars), could allow material to be channelled to the stellar surface via the magnetosphere.
    This is reminiscent of the accretion scheme of DQ Tau and AK Sco \citep[][respectively]{Pouilly23, Pouilly24}, where a magnetospheric accretion is ongoing on the components surrounded by a circumbinary disc too far to be directly truncated by the magnetic field.

    \subsection{Other periodicity detected in accretion structures}

    A tentative periodicity of $f \approx 0.3$ d$^{-1}$ ($\sim$3.3 days) was detected in the Balmer line centres, the H$\gamma$ EW modulation, and the \ion{He}{I} D3 NC (in the primary's velocity reference frame).
    This periodicity indicates that the ESPaDOnS time series covers only two periods.
    Furthermore, the EW of H$\gamma$ in Fig.~\ref{fig:EWbalmer} appears relatively stable, with the exception of a extremum at HJD 2\,455\,081.72.
    This value, likely due to a flare also reported by \cite{Donati11b}, may be responsible for the detection of the 3.3-day period.
    To verify the validity of this period, we first generated a 3.3-day periodic signal (sine wave), sampled at the observation dates, to confirm that the sampling allowed for the detection of this period.
    A GLS periodogram analysis successfully recovered the periodicity, with a FAP of 10$^{-3}$.
    We then performed a leave-one-out test on the H$\gamma$ flux in the velocity channel where the minimum FAP was achieved.
    This involved iteratively removing one observation from the dataset and computing the associated periodogram and FAP.
    All eight periodograms showed a peak between 0.29 and 0.31 d$^{-1}$, with a FAP between 10$^{-1}$ and 10$^{-2}$.
    These tests demonstrate that the 3.3-day periodicity is genuinely detected and not solely induced by the flare at HJD 2\,455\,081.72.

\begin{figure}
        \centering
        \includegraphics[width=0.9\linewidth]{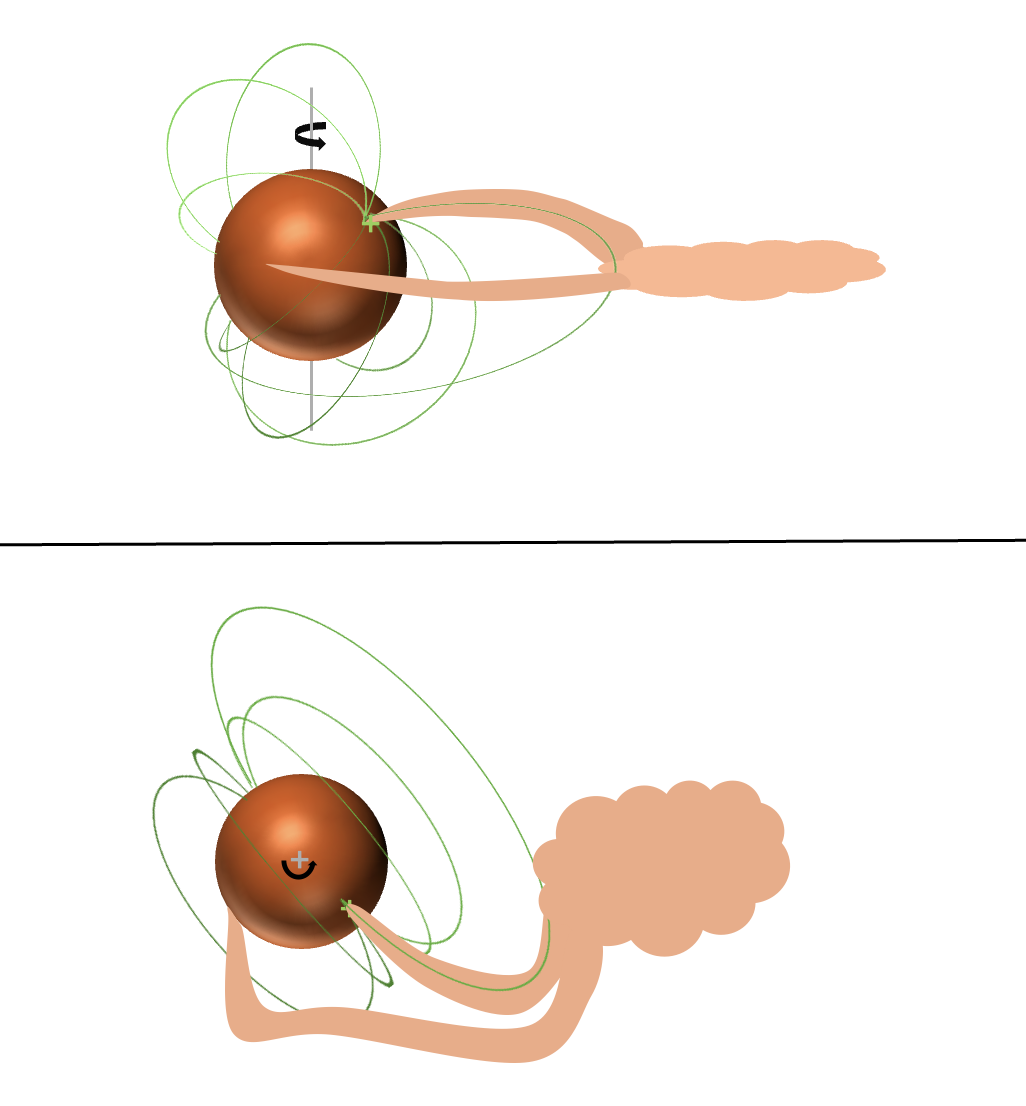}
        \caption{Sketch representing the first hypothesis for the 0.3 d$^{-1}$ period detected where the primary shows both an accretion funnel flow and an accretion tongue. The top sketch shows the side view, the bottom one represents de polar view. The grey line/cross is the rotation axis, the green cross is the dipole pole, the green lines are dipole magnetic field lines. The cloud represents the bulk of gas mentioned in Sect.~\ref{sec:discussion}.}
        \label{fig:schemeHyp1}
    \end{figure}

\begin{figure}
        \centering
        \includegraphics[width=0.9\linewidth]{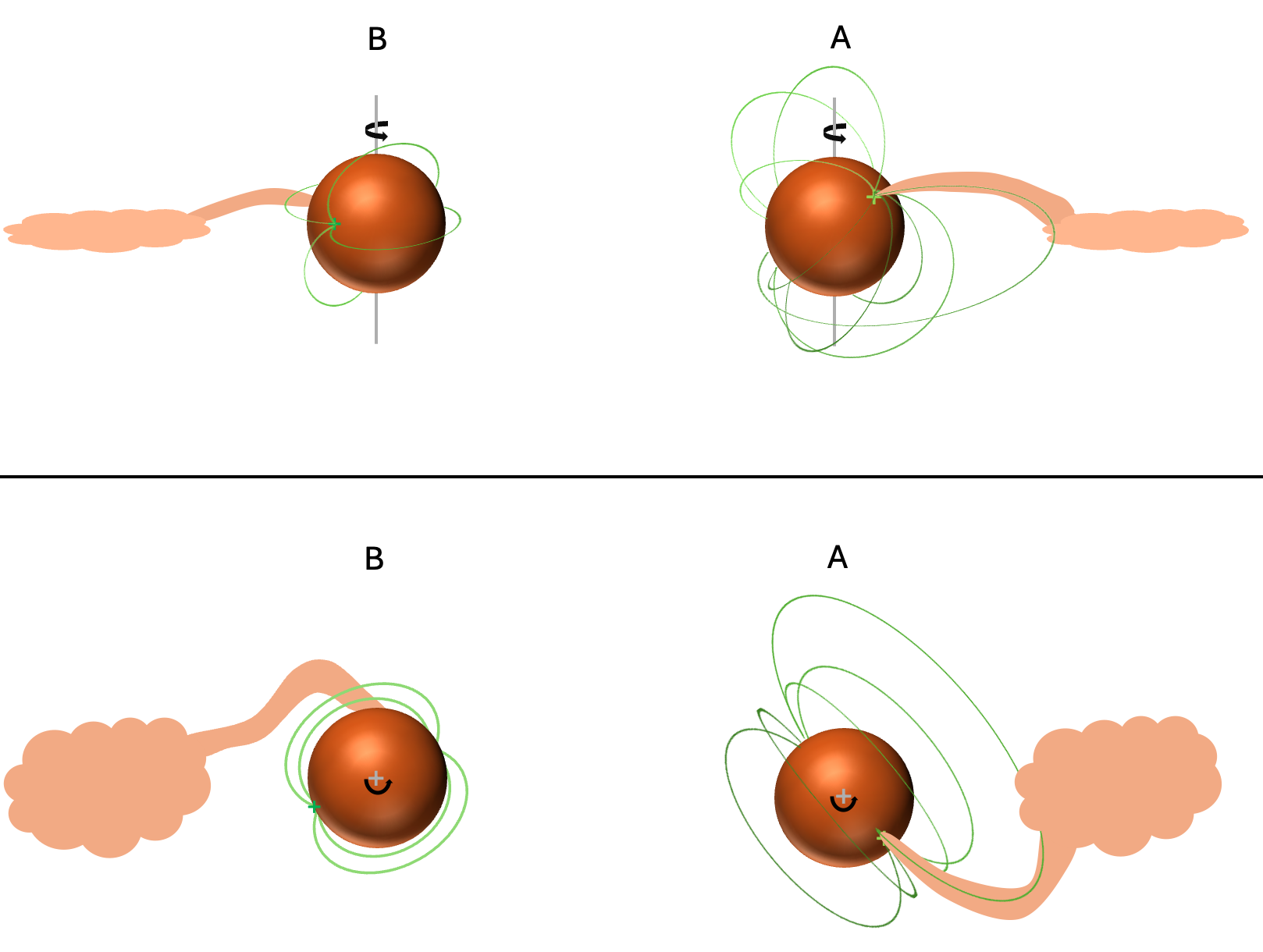}
        \caption{Same as Fig.~\ref{fig:schemeHyp2} for the second hypothesis. The 0.3 d$^{-1}$ period is induced here by the simultaneous passages of the accretion funnel flow of the primary, and an accretion tongue on the secondary.}
        \label{fig:schemeHyp2}
    \end{figure}

    This periodicity likely corresponds to accretion structures cycling through the line of sight but does not match any harmonic of the rotation period.
    We suspect this behaviour to be implied by two different structures, like two accretion funnel flows with two associated accretion shocks, the second appearing on the line of sight $\sim$0.9~d (0.37 rotation cycle) after the first one.
    This would implies a delay of 3.3~d between the the first appearance of the structure \#1 and the second appearance of structure \#2, and between the second passage of structure \#1 and the third of structure \#2, yield the detection of the 3.3-day period.
    We propose two scenarios:
    \begin{itemize}
        \item Accretion tongue on the primary (Fig.~\ref{fig:schemeHyp1}): A secondary accretion tongue penetrates the primary’s magnetosphere to reach the stellar surface at phase 0.66, emitting region of the \ion{Ca}{ii} IRT NC (see Sect.~\ref{subsec:caIIirt}).  Together with the accretion funnel flow at phase 0.37 producing the IPC profile in H$\gamma$ and the minimum EW of the \ion{Ca}{ii} IRT NC, it would imply a 0.29 phase shift between the two structures, only 4.7 min away from the 0.9-day shift aforementioned.
        
        \item Accretion tongue on the secondary (Fig.~\ref{fig:schemeHyp2}): An accretion tongue connect the secondary to its associated bulk of gas, reaching the stellar surface at phase 0.01, emitting region of its \ion{Ca}{ii} IRT NC (see Sect.~\ref{subsec:caIIirt}). The phase shift between such a structure and the accretion funnel flow of V4046 Sgr A is 0.36 very close to the 0.9 d aforementioned, their successive passage could thus implied the 3.3-day period detected.
    \end{itemize}

    Given that the 0.3 d$^{-1}$ periodicity is predominantly detected when using the primary's velocity reference frame, we could favour the first hypothesis.
    However, the accretion is likely originating from the bulk of gas identified by \cite{Stempels04}, implying a single direction of infall towards the stellar surface.
    The first hypothesis thus suggests that the accretion tongue spirals around the star for approximately 0.8$\pi$ radians from the bulk of gas to the stellar surface, and in the direction opposite to the star's rotation.
    While such behaviour is observed in the chaotic accretion regime simulations of \cite{Romanova15}, it is not characteristic of the "ordered chaotic regime", which involves both accretion funnel flows following magnetic field lines and accretion tongues penetrating the magnetosphere.

    Under the second hypothesis, the material spirals for only 0.5$\pi$ radians, still counter-rotating, but as the secondary appears to be in a chaotic accretion regime, we currently favour this explanation.



\section{Conclusions}
\label{sec:conclusion}
    In this paper, we present the analysis of the accretion process in V4046 Sgr, a binary system surrounded by a circumbinary disc and composed of two similar cTTSs in a circularised orbit with synchronised rotation.
    The aim of this study was to build on our previous efforts with DQ Tau and AK Sco to understand the accretion processes in young multiple systems by systematically accounting for the stellar magnetic field.
    Specifically, we sought to explore how the magnetospheric accretion process observed in single cTTSs, where a strong dipolar magnetic field truncates the disc and forces material to leave the disc plane that free-fall onto the stellar surface along magnetic field lines, can be extended to systems with higher multiplicity.

    We analysed an ESPaDOnS spectropolarimetric time series that was previously studied by \cite{Donati11b} and \cite{Hahlin22} in the context of the system's magnetic field.
    Given the pivotal role of magnetic fields in the accretion processes of cTTSs, these prior works provide a foundation for our complete and detailed characterisation of the accretion process in V4046 Sgr.

    Our findings reveal an uneven accretion process between the two components of the system, originating from bulks of gas located near the co-linear Lagrangian points, rather than directly from the circumbinary disc.
    V4046 Sgr A displays clear signatures of magnetospheric accretion, primarily inferred from the IPC profiles detected in the Balmer lines and from the variations in the primary’s NC of the \ion{Ca}{II} IRT line.
    These features trace an accretion funnel flow and accretion shock in phase with the primary’s dipole pole.

    In contrast, the secondary component does not exhibit such behaviour.
    However, variations in the Balmer lines and the \ion{He}{I} D3 line indicate that this component is the main accretor of the system.

    The extensive emitting regions of the NCs of the \ion{Ca}{II} IRT lines for both components suggest that accretion is occurring across a wide area of the stellar surfaces, rather than being confined to the concentrated locations expected for typical magnetospheric accretion processes.
    Consequently, we conclude that the primary operates under an "ordered chaotic" accretion regime, where accretion funnel flows (connecting the disc to the star’s dipole pole) coexist with accretion tongues (penetrating the magnetosphere to reach the stellar surface near the equator).

    For the secondary, we suspect a chaotic accretion regime, characterised exclusively by accretion tongues, consistent with its highly inclined dipole field.


\begin{acknowledgements}

    We thanks the anonymous referee, whose comments help in improving the strength of our work.
    
    This research was funded in whole or in part by the Swiss National Science Foundation (SNSF), grant number 217195 (SIMBA).
    
    Based on observations obtained at the Canada–France– Hawaii Telescope (CFHT) which is operated from the summit of Maunakea by the National Research Council of Canada, the institut National des Sciences de l’Univers of the Centre National de la Recherche Scientifique of France, and the University of Hawaii. The observations at the Canada–France–Hawaii Telescope were performed with care and respect from the summit of Maunakea which is a significant cultural and historic site.

    This work has made use of the VALD database, operated at Uppsala University, the Institute of Astronomy RAS in Moscow, and the University of Vienna.

    The \texttt{SpecpolFlow} package is available at \url{https://github.com/folsomcp/specpolFlow}

    The \texttt{PySTEL(L)A} package is available at \url{https://github.com/pouillyk/PySTELLA}.

\end{acknowledgements}

\bibliographystyle{aa}
\bibliography{literature}

\begin{appendix}
    \section{Radial velocities}
    \label{ap:vr}
        Here we provide the radial velocities derived for each components and for each observation using our disentangling method of LSD Stokes \textit{I} profiles. This method is described in Sect.~\ref{subsec:LSDVr} and the curve is shown in Fig.~\ref{fig:vradLSD}.

        \begin{table}[H]
        \centering
        \caption{Radial velocities obtained from the LSD Stokes \textit{I} disentangling procedure (see Sect.~\ref{subsec:LSDVr}).}
        \begin{tabular}{lll}
            \hline
            HJD & \vr(A)  &  \vr(B) \\
            ($-$2\,450\,000 d) & (\kms) & (\kms) \\
            \hline
            5077.774410 & 17.502$\pm$0.095 & $-$30.48$\pm$0.14 \\
            5078.817320 & $-$7.79$\pm$0.13 & $-$2.51$\pm$0.15 \\
            5079.722230 & $-$41.272$\pm$0.094 & 32.12$\pm$0.17 \\
            5080.722140 & 44.311$\pm$0.086 & $-$58.44$\pm$0.14 \\
            5080.814130 & 39.782$\pm$0.095 & $-$53.56$\pm$0.13 \\
            5081.722050 & $-$55.56$\pm$0.11 & 46.88$\pm$0.12 \\
            5082.722960 & 29.017$\pm$0.091 & $-$42.71$\pm$0.16 \\
            5083.721860 & $-$15.05$\pm$0.11 & 4.99$\pm$0.15 \\
            \hline
        \end{tabular}
        \end{table}

    \section{Equivalent widths}
    \label{ap:ew}
        In this appendix, we present a table summarising the EWs described in Sect.~\ref{sec:results}.

        \begin{table}[H]
        \centering
        \caption{EWs of lines computed in this work.}
        \begin{tabular}{llllll}
            \hline
            & \multicolumn{5}{c}{EW} \\
            \cline{2-6}
            HJD & H$\alpha$ & H$\beta$ & H$\gamma$ & \ion{Ca}{II} IRT NC (A) & \ion{Ca}{II} IRT NC (B) \\
            ($-$2\,450\,000 d) & (nm) & (nm) & (nm) & (nm) & (nm)\\
            \hline
            5077.774410 & $-$5.72$\pm$0.07 & $-$0.90$\pm$0.09 & $-$0.43$\pm$0.11 & $-$0.025$\pm$0.002 & $-$0.023$\pm$0.002 \\
            5078.817320 & $-$6.56$\pm$0.06 & $-$0.91$\pm$0.09 & $-$0.45$\pm$0.11 & $-$0.020$\pm$0.003 & $-$0.018$\pm$0.003 \\
            5079.722230 & $-$6.54$\pm$0.06 & $-$1.02$\pm$0.08 & $-$0.48$\pm$0.10 & $-$0.020$\pm$0.002 & $-$0.016$\pm$0.002 \\
            5080.722140 & $-$6.13$\pm$0.06 & $-$0.66$\pm$0.08 & $-$0.36$\pm$0.10 & $-$0.020$\pm$0.002 & $-$0.018$\pm$0.002 \\
            5080.814130 & $-$6.16$\pm$0.06 & $-$0.70$\pm$0.08 & $-$0.34$\pm$0.10 & $-$0.019$\pm$0.002 & $-$0.017$\pm$0.002 \\
            5081.722050 & $-$8.18$\pm$0.07 & $-$1.64$\pm$0.09 & $-$0.85$\pm$0.11 & $-$0.021$\pm$0.003 & $-$0.018$\pm$0.002 \\
            5082.722960 & $-$5.29$\pm$0.06 & $-$0.73$\pm$0.08 & $-$0.32$\pm$0.10 & $-$0.018$\pm$0.002 & $-$0.016$\pm$0.002 \\
            5083.721860 & $-$4.21$\pm$0.05 & $-$0.41$\pm$0.08 & $-$0.22$\pm$0.09 & $-$0.023$\pm$0.002 & $-$0.020$\pm$0.002 \\
            \hline
        \end{tabular}
        \tablefoot{Velocity ranges used for the derivation: H$\alpha$: [$-$400;$+$500]~\kms, H$\beta$: [$-$350;$+$450]~\kms, H$\gamma$: [$-$350;$+$450]~\kms, \ion{Ca}{II} IRT NCs: [$-$50;$+$50]~\kms.}

        \end{table}

\FloatBarrier
    \section{TESS light curves folded in phase}
    In this appendix, we provide another visualisation of the TESS light curves shown in Fig.~\ref{fig:TESSlc}. 
    Here only the curves folded in phase are presented, but each cycle has been shifted vertically to better appreciate the individual variabilities.
    
        \begin{figure}
            \centering
            \includegraphics[width=0.49\linewidth]{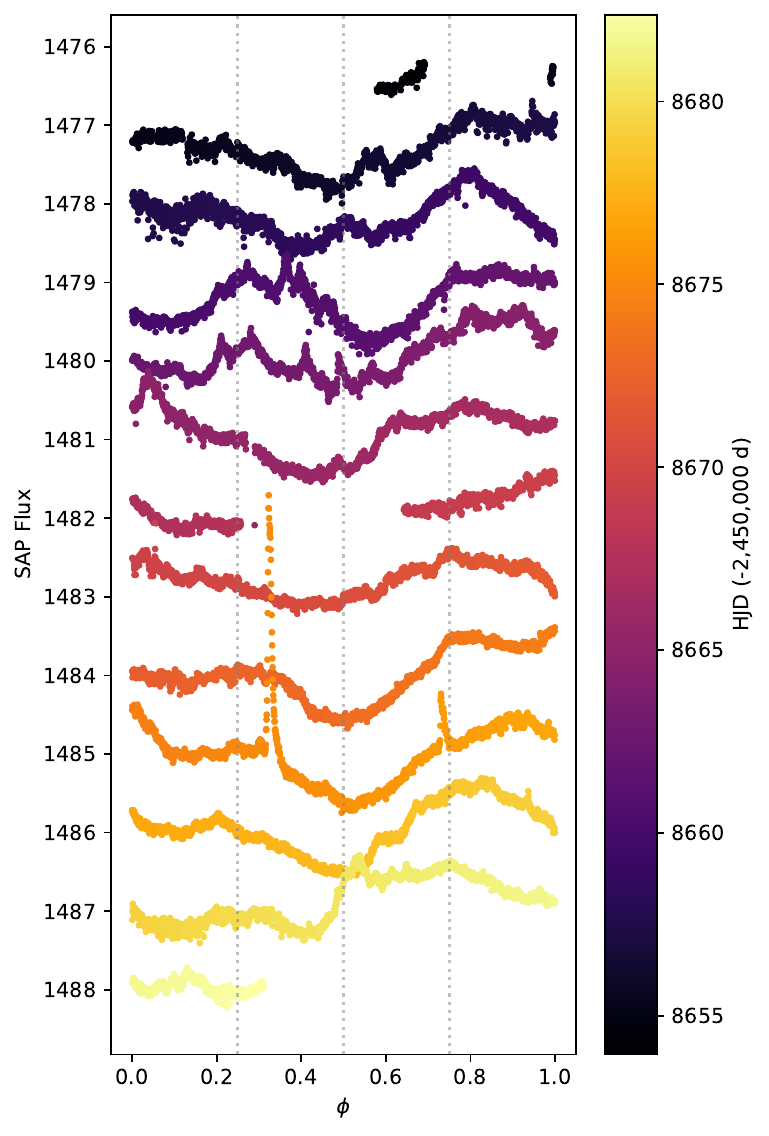}
            \includegraphics[width=0.49\linewidth]{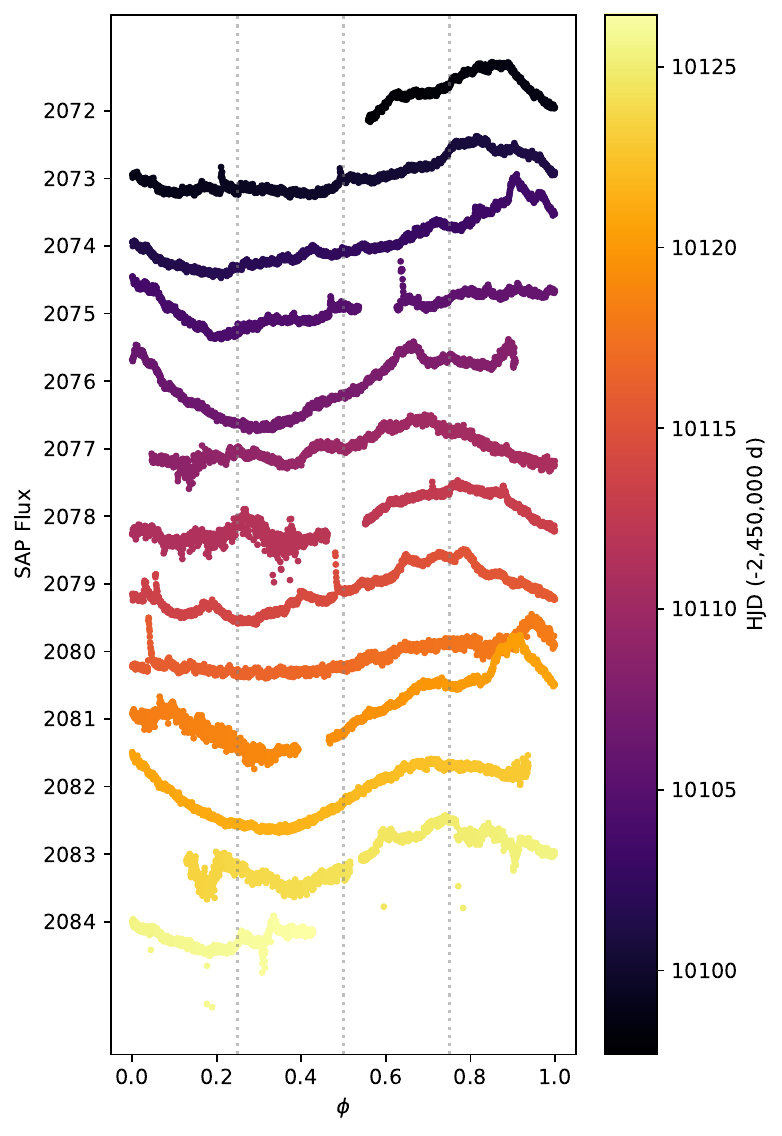}
            \caption{Sector 13 \textit{(left)} and 66 \textit{(right)} TESS light curves folded in phase following Eq.~\ref{eq:ephemeris}. The different cycles have been shifted vertically to improve the readability. The vertical grey dotted lines illustrate the phase 0.25 (the A component is in front of the observer), 0.5, and 0.75 (the B component is facing the observer). The y axis labels denote the orbital cycle from the first ESPaDOnS observation used in this work.}
            \label{fig:phaseTESSlc}
        \end{figure}

\end{appendix}

\end{document}